\documentclass[useAMS,referee,usenatbib]{biom} 

\usepackage{natbib}
\usepackage{bbold}
\usepackage{multirow} 
\usepackage{pdflscape}
\usepackage{setspace}
\usepackage{graphicx}
\usepackage{algorithm}
\usepackage{algpseudocode}
\usepackage{tabularx}
\usepackage{booktabs}
\usepackage{ragged2e}
\usepackage{blindtext}
\usepackage{amssymb,amsmath}
\usepackage{caption}
\usepackage{bbm}
\usepackage{mathrsfs}
\usepackage[table]{xcolor}

\newcommand{\indep}{\perp \!\!\! \perp}
\usepackage[skip=0pt]{parskip}
\title[]{Doubly Robust Targeted Estimation of Subgroup Average Treatment Effects for Time-to-event Outcomes with Competing Risks}

\author{Runjia Li$^{1}$, 
Victor B. Talisa$^{2}$, and Chung-Chou H. Chang$^{1,3,*}$ \email{changj@pitt.edu} \\
$^{1}$Department of Biostatistics, University of Pittsburgh, Pittsburgh, Pennsylvania, U.S.A. \\
$^{2}$Department of Critical Care, University of Pittsburgh, Pittsburgh, Pennsylvania, U.S.A. \\
$^{3}$Department of Medicine, University of Pittsburgh, Pittsburgh, Pennsylvania, U.S.A. }

\begin{document}
\label{firstpage}

\begin{abstract}
In recent years, precision treatment strategies have gained significant attention in medical research, particularly for patient care. We propose a novel framework for estimating subgroup average treatment effects (ATE) in time-to-event data with competing risks, using ICU patients with sepsis as an illustrative example. Our approach, based on cumulative incidence functions and targeted maximum likelihood estimation (TMLE), achieves both asymptotic efficiency and double robustness. The primary contribution of this work lies in our derivation of the efficient influence function for the targeted causal parameter, subgroup ATE. We established the theoretical proofs for these properties, and subsequently confirmed them through simulations. Our TMLE framework is flexible, accommodating various regression and machine learning models, making it applicable in diverse scenarios. In order to identify variables contributing to treatment effect heterogeneity and to facilitate accurate estimation of subgroup ATE, we developed two distinct variable importance measures (VIMs). This work provides a powerful tool for optimizing personalized treatment strategies, furthering the pursuit of precision medicine.
\end{abstract}

\maketitle


\section{Introduction}

Precision medicine requires methods that account for patient heterogeneity, particularly in life-threatening conditions such as sepsis in the intensive care unit (ICU), where outcomes are time-to-event with competing risks. In this context, robust methods are needed to estimate subgroup average treatment effects (ATEs) to inform tailored interventions.

The subdistribution hazards model by \cite{fine1999proportional} is the most widely used method for analyzing time-to-event data in the presence of competing risks. To avoid issues of non-collapsibility \citep{gail1984biased}, we focus on the estimand of differences in the cumulative incidence function (CIF) to infer causality. Unlike approaches based on cause-specific hazards of all competing events, the subdistribution hazards model directly connects covariate effects to the CIF, simplifying interpretation. Thus, our targeted estimand is a natural choice for estimating patient prognosis and predicting CIFs \citep{koller2012competing, austin_practical_2017}.

Doubly robust estimators provide consistent estimates when either the outcome or treatment model is correctly specified and achieve efficiency when both are correct \citep{robins1994estimation, van_der_laan_targeted_2006}. Among these, targeted maximum likelihood estimation (TMLE) \citep{van_der_laan_targeted_2006} is particularly advantageous in finite samples as a substitution estimator, ensuring parameter bounds, which is critical when the estimand is defined by probabilities. \citet{wei2023efficient} further developed a one-step TMLE for binary and continuous outcomes that efficiently evaluates heterogeneous treatment effects across multiple subgroups simultaneously and remains robust to small estimated propensity scores. TMLE has also been extended to survival outcomes, with additional work for censoring adjustment \citep{moore2009application,cai2020one,benkeser_improved_2018,rytgaard2023estimation,rytgaard2024targeted}. However, while some TMLE-based studies have aimed at CIF estimation, they often rely on cause-specific hazard functions, where covariate effects do not directly reflect effects on the CIF, and they generally focus on average treatment effects, providing limited guidance for subgroup-level treatment decisions. 

In this study, we propose a comprehensive TMLE framework to consistently estimate subgroup ATEs in time-to-event data with competing risks. The primary contribution is our derivation of the efficient influence function (EIF) for the target causal parameter, subgroup ATE, defined by difference in potential CIFs under distinct treatment conditions and baseline characteristics. Our estimator possesses double robustness, achieves asymptotic efficiency under mild conditions, and accommodates flexible machine learning algorithms for nuisance estimation. Leveraging the empirical variance of the EIF, we construct confidence intervals and conduct hypothesis tests to detect overall treatment effects and effect heterogeneity. In addition, we develop two complementary variable importance measures (VIMs) to quantify the contribution of each covariate to treatment effect heterogensity: one measures predictive importance and the other measures prognostic importance. These VIMs provide interpretable metrics to prioritize variables for precision medicine applications.

This paper is organized as follows. Section \ref{c2s:notation} introduces the notation for competing risk data and defines the causal parameters of interest. Section \ref{c2s:method} presents our proposed TMLE framework and the properties of the proposed estimator. Section \ref{c2s:sim} reports simulation studies evaluating performance in multiple scenarios. Section \ref{c2s:app} applies the method to ICU sepsis data to estimate subgroup ATEs of steroids. Section \ref{c2s:dis} concludes with a discussion.

\vspace{-2pt}
\section{Setting and Notation} \label{c2s:notation}
\subsection{Time-to-event data with competing risks}\label{c2ss:setting}
We consider a time-to-event data setting with competing risks and two treatment arms. Let the observed data be denoted as $O_i=\left(X_i, A_i, \widetilde{T}_i, \widetilde{\Delta}_i\right)_{i=1}^N\sim^{i.i.d }P$ for subjects $i=1, \ldots, N$, where $X\in \mathcal{X}\subseteq \mathbb{R}^p$ represents a vector of baseline covariates, $A\in\{0,1\}$ is a binary treatment assignment, $T$ is the failure time, and $C$ is the right-censoring time. $P$ denotes the unknown true probability distribution of $O$, from which our observed sample is drawn. Define $\widetilde{T}=T\wedge C$ as the observed event time, $\Delta\in\{1,2\}$ as the event type, and $\widetilde{\Delta}=\mathbbm{1}(T\leq C)\Delta$ as the observed event type. Without loss of generality, let event 1 ($\Delta=1$) be the primary event of interest, and event 2 ($\Delta=2$) be the competing event. The cumulative incidence function (CIF) for the primary event at time $t_0$ is defined as $F_1(t_0)=\operatorname{Pr}(T\leq t_0, \Delta=1)$. Let $T^a$ and $\Delta^a$ be the potential event time and event type, respectively, had the subject been assigned treatment $A=a$ $(a=0,1)$. The potential CIF for the primary event at time $t_0$ had the subject received treatment $a$ is $F_1^a(t_0)=\operatorname{Pr}(T^a\leq t_0, \Delta^a=1)$.
\vspace{-2pt}
\subsection{Causal estimand}\label{c2ss:causalest}
Our parameter of interest is the subgroup average treatment effect (ATE) in prespecified subgroups by a set of predictive variables $V\in \mathcal{V}\subset \mathcal{X}$, where treatment effects differ among patients with various values of $V$. Variables $V$ are usually selected based on clinical knowledge, interests, or by a data-driven approach. $L\in \mathcal{L}\subseteq \mathcal{X}\backslash \mathcal{V}$ denotes prognostic variables that affect CIF but are not main contributors to the heterogeneity of the treatment effect.
Let \(\mathcal{V}_m\subseteq\mathcal{V}\) denote the \(m\)th prespecified subgroup region, \(m=1,2,\ldots,M\). The subgroup average treatment effect conditional on $V\in \mathcal{V}_m$  is defined as the difference in potential CIF of main event:
$$
\begin{aligned}
\Psi_{t_0,m}(P)&=F^1_{1}(t_0|V\in \mathcal{V}_m)-F^0_{1}(t_0|V\in \mathcal{V}_m)\\
&=E_L\left\{F^1_{1}(t_0|V\in \mathcal{V}_m,L)-F^0_{1}(t_0|V\in \mathcal{V}_m,L)\right\}
\end{aligned}
$$
where $F^1_{1}(t_0|V\in \mathcal{V}_m,L)$ is the potential CIF conditional on the prognostic variables $L$ and  predictive variables $V\in \mathcal{V}_m$. When $\mathcal{V}=\varnothing$, $M=1$ and $\Psi_{t_0}(P)$ is the ATE.

Within the potential outcome framework by \cite{rosenbaum1983central}, this parameter can be causally identified from observed data under the following assumptions: 1) Consistency: $T=T^a$ and $\Delta =\Delta^a$ when the assigned treatment A=a; 2) No unmeasured confounding: $T^a,\Delta^a \indep {A}| V,L$; 3) Positivity: $0<\operatorname{Pr}(A=1|V,L)<1 $ and $\operatorname{Pr}(C>t_0|A,V,L)>0$ with probability 1; and 4) Coarsening at random censoring: \(T,\Delta \perp C \mid A,V,L\), meaning that, conditional on treatment and observed covariates, censoring does not depend on the unobserved event time or type. This is the standard CAR condition for right-censored data   \citep{gill1997coarsening,tsiatis2006semiparametric}. In addition, we assume that the potential outcomes are not affected by the treatment assignments of other subjects, and only one version of the treatment (known as the stable unit treatment value assumption, SUTVA). Under these assumptions, the target causal parameter within each subgroup can then be identified as:
\begin{equation}\label{sATE0} 
\begin{aligned}
\Psi_{t_0,m}(P)&=E_L\left\{F_1(t_0|A=1,L,V\in \mathcal{V}_m)-F_1(t_0|A=0,L,V\in \mathcal{V}_m)\right\}\\
&=E_L\left\{F_{1,m}(t_0|A=1,L)-F_{1,m}(t_0|A=0,L)\right\},
\end{aligned}
\end{equation} 
where $F_{1,m}(t_0 \mid A = a, L) \equiv F_1(t_0 \mid A = a, L, V\in \mathcal{V}_m)$ denotes the potential CIF for cause 1 at time $t_0$ in subgroup $V\in \mathcal{V}_m$ under treatment $A = a$.


\vspace{-3pt}
\section{Targeted MLE Framework} \label{c2s:method}
We developed a doubly robust estimation framework to construct the targeted estimator $\Psi_{t_0,m}(\hat P_n^*)$. This framework is based on the targeted maximum likelihood estimation (TMLE) introduced by \cite{van_der_laan_targeted_2006}, along with methods extended to estimating average treatment effect for time-to-event outcomes \citep{rytgaard2024targeted, rytgaard2023estimation,moore2009application,cai2020one}. When subjects are stratified into subgroups according to a predetermined set of predictive covariates, our estimation procedure is applied \textbf{within each subgroup} follows these steps: 1) \textbf{Initial estimation}: we begin by generating initial estimates for the nuisance parameters; 2) \textbf{Targeting step}: the estimators are updated iteratively to solve the corresponding efficient influence function (EIF) equations, thereby ensuring asymptotic properties; and 3) \textbf{subgroup ATE estimation}: finally, we compute the subgroup ATE by averaging the updated estimates on the baseline covariates that are not used for stratification.  


In the remainder of this section, we drop the subscripts $m$ and $t_0$ for notational simplicity, with the understanding that the procedure is carried out separately for each subgroup. Our approach targets estimation and pointwise inference within each subgroup $V\in \mathcal{V}_m$, rather than uniform inference across all subgroups. Further methodological details are provided in the following subsections, with a step-by-step summary in Web Appendix A (Algorithm 1).
\vspace{-2pt}
\subsection{Initial estimation} \label{c2ss:initial}
To solve the EIF equations in the targeting step, 
we need initial estimates for the nuisances: (i) CIF of the main event $F_{1}(t|A,L)=1-\exp\{-\int_0^t\lambda_1(s|A,L)ds\}$, where $\lambda_1(s|A,L)$ is the subdistribution hazard function of the main event; (ii) survival function for censoring $G(t|A,L)=\operatorname{Pr}(C>t|A,L)$; and (iii) the treatment mechanism $\pi(a|L)=\operatorname{Pr}(A=a|L)$.

To estimate the CIF, let's start with the subdistribution hazard function \citep{gray_class_1988}
\begin{equation}\label{e:subdist hazard}
\begin{aligned}
\lambda_1\left(t| A, L\right) =\lim _{dt \rightarrow 0} \frac{1}{dt} \operatorname{Pr}&\{t \leq T < t+dt, {\Delta}=1 \mid \\ &T \geq t \cup(  T \leq t \cap{\Delta} \neq 1), A, L\}, 
\end{aligned}
\end{equation}
where the risk set includes subjects who are event-free and those who have previously experienced a competing event. Because there is a one-to-one relationship between the subdistribution hazard function and the CIF, we use the subdistribution hazards model proposed by \cite{fine1999proportional} to predict the CIF. To simultaneously estimate unknown parameters and perform variable selection, \cite{fu_penalized_2017} developed penalized subdistribution hazards model and related algorithms. \cite{kawaguchi_fast_2020} developed a computationally efficient algorithm for fitting these penalized models in R package \texttt{fastcmprsk}. We adopted their algorithm in our parameter estimation and variable selection procedure.

The conditional survival function for censoring events can be estimated using various methods, including the Kaplan-Meier estimator or regression models like the Cox proportional hazards model. Similarly, the treatment mechanism can be estimated using a range of statistical techniques for binary outcome, such as logistic regression and generalized additive models (GAM) \citep{van2011targeted}.

\subsection{Targeting step: iterative updates}\label{c2ss:update}
To achieve asymptotic efficiency and doubly robust properties for the estimator, the key step in TMLE involves refining the initial estimates from Step 1 by deriving the EIF expression and solving the estimating equation. We will first explain the underlying theory and then illustrate the process of obtaining the asymptotically efficient and doubly robust estimator.

\vspace{-5pt}
\subsubsection{Efficient influence function (EIF)} \label{c2sss:EIF}
The EIF has been widely used in constructing efficient and doubly robust estimators \citep{robins1994estimation,van_der_laan_targeted_2006} based on the semiparametric and empirical process theories \citep{bickel1993efficient,van2000asymptotic}. Under regularity conditions, the targeted estimator $\Psi(\hat P_n^*)$ is asymptotically linear if 
\begin{equation} \label{e:asym linear}
\sqrt{n}\{\Psi(\hat P_n^*)-\Psi(P)\}=\sqrt{n} \mathbb{P}_n\{D(P)\}+o_P(1),
\end{equation}
where $D(P)$ is an influence function of the estimator, $\mathbb{P}_n$ denote the empirical average that $\mathbb{P}_n(\cdot)=\frac{1}{n}\sum_{i=1}^n(\cdot)$,  $E\{D(P)\}=0$, and $E\{D(P)^2\}<\infty$. A regular estimator in a semiparametric model is asymptotically efficient with the lowest asymptotic variance if it is asymptotically linear with $D(P)$ being the EIF \citep{bickel1993efficient}. 
In this work, the target parameter is viewed as a statistical functional \(\Psi :\mathcal{M}\rightarrow\mathbb{R}\) in a nonparametric model \(\mathcal{M}\), so the EIF corresponds to the only influence function $D(P)$ \citep{bickel1993efficient}.

To obtain the asymptotically efficient estimator and calculate the lower bound of the variance, we need to: 1) derive the expression of $D(P)$, and 2) find the targeted distribution $\hat P_n^*=(\hat\lambda_1^*, \hat \pi, \hat G)$ that solves the efficient influence function equation:
\begin{equation}\label{EICequation}
\mathbb{P}_n\{D(\hat P_n^*)\}=0,
\end{equation}
where $\hat \lambda_1^*$ is the updated estimated subdistribution hazard function.

 A general approach for deriving the EIF is to calculate the pathwise derivative $\frac{d}{d\epsilon} \Psi(P_\epsilon)\Bigr|_{\epsilon=0}$ for a submodel $P_\epsilon$. Unlike a conventional parametric model, a submodel is a useful conceptual tool in semiparametric theory, instead of being used for analysis \citep{tsiatis2006semiparametric}. 
A commonly chosen submodel has the form: $dP_\epsilon=(1+\epsilon b)dP$ \citep{bickel1993efficient, van2011targeted}, where  $\epsilon\in \mathbb{R}$ indexes the perturbation; $b=b( T, \Delta, A, L)$ is a zero-mean score function and for this submodel, $b=\frac{d}{d\epsilon} \log(dP_\epsilon)\Bigr|_{\epsilon=0}$. The EIF can be obtained by expressing the pathwise derivative of the target estimand as an inner product of two terms:
 $\frac{d}{d\epsilon} \Psi(P_\epsilon)\Bigr|_{\epsilon=0}=E\{D(P)\cdot b\}$ \citep{levy2019tutorial}.
By definition of the CIF, our target parameter can be expressed as:
\begin{equation} \label{functional}
\begin{aligned}
\Psi(P) &=E\left\{F_1(t_0|A=1,L)-F_1(t_0|A=0,L)\right\}\\
&=\int_{\mathcal{L}} \sum_{a=0,1}\left[1-\exp \left\{-\int_{0}^{t_{0}} \Lambda_{1}(d t | a,l)\right\}\right] \left\{\mathbbm{1}(a=1)-\mathbbm{1}(a=0)\right\} \mu(l) dl,
\end{aligned}
\end{equation}
where $\mu(l)$ is the density of prognostic variables $L$.

Let $P^{\delta}(d t | a,l)=\operatorname{Pr}(t\leq{T} <t+dt, \Delta=\delta|a,l)$, then from the definition of subdistribution hazard function (\ref{e:subdist hazard}) we have $\Lambda_{1, \epsilon}(d t | a, l)=P_\epsilon^1(dt | a,l)\{P_\epsilon({T} \geq t	\cup ({T} \leq t \cap {\Delta} \neq 1) | a, l)\}^{-1}$. The derivation relies on the differentiation of $P^\delta_{\epsilon}$ with respect to $\epsilon$ evaluated at $\epsilon=0$, i.e., $\frac{d}{d \epsilon} P^\delta_{\epsilon}(d t | a, l) \Bigr| _{\epsilon=0}=\frac{d}{d \epsilon} P^\delta(d t | a, l)(1+\epsilon b) \Bigr| _{\epsilon=0} =P^\delta(d t | a, l) b(t,\delta,a,l)$.   \vspace{2em} \\
\textbf{Theorem 1.}\label{c2thm:eif} \textit{Efficient influence function of $\Psi(P)$}. \normalfont Under Assumptions 1 - 4 in Section \ref{c2ss:causalest}, the efficient influence function has the form of:
\begin{equation} \label{EIF}
\begin{aligned} 
 D(P)=&\int_0^{t_0} \left\{h(t,A,L)w(t) M(d t | A, L)\right\} +\left\{F_{1}(t_{0} | 1,L)-F_{1}(t_{0} | 0,L)\right\}-\Psi(P),
\end{aligned}
\end{equation}
where \begin{equation}\label{clever}
\begin{aligned}
    h(t,A,L)=&\frac{\{\mathbbm{1}(A=1)-\mathbbm{1}(A=0)\}}{\pi(A| L) G(t^-| A, L)} \cdot \frac{1-F_{1}\left(t_{0} | A, L\right)}{1-F_1 (t | A, L)}
\end{aligned}
\end{equation} 
is defined as the \textit{clever covariate}, which plays a pivotal role in the fluctuation model for estimating subgroup ATE under the TMLE framework (will be described in detail in next subsection); $w(t)=\mathbbm{1}(C\geq T\wedge t)G(t^-| A,L)/G\{(\tilde T\wedge t)^-| A,L\}$ is the weight determined by adapting the inverse probability of censoring weighting (IPCW) technique \citep{fine1999proportional}; and $M(dt| A,L)=N(dt)-Y(t)\Lambda(dt| A,L)$, where $N(t)=\mathbbm{1}(T\leq t,\Delta=1)$ and $Y(t)=\{1-N(t^-)\}$. Derivation of EIF is in Web Appendix B. 
\subsubsection{Solving the efficient influence function equation}\label{c2sss:submodel}
The targeted estimator $\Psi(\hat P_n^*)$ is obtained by solving the efficient influence function equation (\ref{EICequation}) via a fluctuation model. The fluctuation model is selected to have its corresponding mean score function equal to the empirical mean of first part of the EIF expressed in (\ref{EIF}) when the fluctuation parameter $\epsilon=0$. In our case, we consider the following fluctuation model: 
\begin{equation}\label{submodel_equa}
\lambda_{1, \epsilon}\left(t | A, L\right) =\lambda_{1}\left(t | A, L\right) \exp \left\{\epsilon h\left(t, A, L\right)\right\},
\end{equation}
where $h\left(t, A, L\right)$ is the clever covariate defined in (\ref{clever}), and $\epsilon \in \mathbb{R}$ is the fluctuation parameter. The score function with respect to this model is: 
\begin{equation}\label{score}
\begin{aligned}
\frac{d \log \mathscr{L} \left(\lambda_{1, \epsilon}\right)}{d \epsilon}
&=\sum_{i=1}^n \int_{0}^{t_0} w_i(t)h(t,A_i,L_i) \left[N_i(d t)-Y_i(t) \lambda_{1}(t|A_i,L_i)\exp\{\epsilon h(t,A_i,L_i)\} d t\right],
\end{aligned}
\end{equation}
where $\mathscr{L}$ is the likelihood function of the model. We obtain the maximum likelihood estimator $\hat \epsilon ^{(k)}$ by solving  $\frac{d \log \mathscr{L}\left(\lambda_{1, \epsilon}\right)}{d \epsilon}=0$. At each iteration (index $k=0,1,2,...$), we use $\hat \epsilon ^{(k)}$ to update $\hat P_n$, specifically, $\hat \lambda_{1}^{(k+1)}(t | A, L)=\hat \lambda_{1}^{(k)}(t | A, L)\exp\left\{\hat \epsilon ^{(k)} h(t,A,L)\right\}$, and $\hat{F}_{1}^{(k+1)}\left(t | A, L\right)=1-\exp \left\{-\int_0^{t} \hat{\lambda}_{1}^{(k+1)}\left(t | A, L\right) d t\right\}$. It will iterate until $\hat \epsilon$ becomes very small and approaches 0. Consequently, the mean score function (\ref{score}) converges to the empirical mean of the first part of the EIF (\ref{EIF}): 
\begin{equation}\label{equal}
    \frac{1}{n}\frac{d \log \mathscr{L} \left(\lambda_{1, \epsilon}\right)}{d \epsilon}\Bigr|_{\epsilon=0}=\mathbb{P}_n\left[\int_0^{t_0} \left\{h(t,A,L)w(t) M(d t | A, L)\right\}\right].
\end{equation}
\setlength{\emergencystretch}{0.5\textwidth}\par
Note that the empirical mean of the remaining part of the EIF is zero by definition, $\mathbb{P}_n[\{\hat F_{1}^*(t_{0} | 1,L)-\hat F_{1}^*(t_{0}|0,L)\}-\Psi(\hat P_n^*)]=0$. The property of (\ref{equal}) enables us to solve the EIF equation (\ref{EICequation}) around $\epsilon=0$ and obtain the solution $\hat P_n^*$. For implementation, we can set a small number $s_n=o_p(n^{-\frac{1}{2}})$ and stop the iteration when $\left |\epsilon\right |\leq s_n $. Derivation of the score function can be found in Web Appendix C. Subsequently, we can substitute this solution to obtain the targeted estimator $\Psi(\hat P_n^*)$, which is introduced in the next subsection. 
The fluctuation model above targets a prespecified \(t_0\); universal least favorable submodels may be preferable for time-indexed targets and are left for future work.

\vspace{-12pt}
\subsection{Asymptotic properties} \label{c2ss:properties}
With the estimator $\hat P_n^*$ being the solution of the EIF equation, $\Psi(\hat P_n^*)$ is asymptotically linear with respect to $D(P)$ if the following conditions hold: 1) the second-order remainder $R(\hat P_n^*,P)=\Psi(\hat P_n^*)-\Psi(P)+E_{P}\{D(\hat P_n^*)\}$ is $o_p(n^{-1/2}$); and 2) $D(\hat P_n^*)$ belongs to the Donsker class \citep{van2000asymptotic, van_der_laan_targeted_2006,van2011targeted}. 

For the first condition, the second-order remainder can be decomposed as
\begin{equation}\label{remainderTerm}
\begin{aligned}
R(\hat P_n^*,P)=\int_{\mathcal{L}}& \sum_{a=0,1} \int_0^{t_0}\left[\frac{\pi(a| l)}{\hat \pi(a| l)} \frac{G\{(\tilde T\wedge t)^-| a,l\}}{\hat G\{(\tilde T\wedge t)^-| a,l\}}-1\right]\cdot\left\{\Lambda_{1}(d t | a, l)-\hat \Lambda_1^*(d t | a, l)\right\}\\ & \cdot \left\{\frac{1-\hat F_1^*(t_0 | a,l)} {1-\hat F_1^*(t | a,l)}\right\} \left\{1-F_{1}(t^-| a, l)\right\}
\cdot\left\{\mathbbm{1}(a=1)-\mathbbm{1}(a=0)\right\}\mu(l)dl,\\
\end{aligned}
\end{equation}
where the double-robust structure is demonstrated. The remainder equals $o_p(n^{-1/2})$ when either 1) both the treatment and censoring are consistently estimated or 2) the CIF is consistently estimated, i.e., either $||\{\pi(a| l)G(t| a,l)\}-\{\hat \pi(a| l){\hat G(t | a, l)}\}|| = o_p(n^{-1/2})$, or $||\Lambda_{1}(dt | a,l)-\hat \Lambda_1^*(dt | a,l)||= o_p(n^{-1/2})$. $R(\hat P_n^*,P)=o_p(n^{-1/2})$ is achieved if $\hat \pi_0(a| l)\hat G(t | a, l)$ and $\hat \lambda_1^*(t|a,l)$ converge faster than $o_P(n^{-1/4})$. See Web Appendix D for details.
For the second condition, we assume that nuisances \(F_1\), \(\lambda_1\), \(G\), and \(\pi\), are c\`adl\`ag functions with finite sectional variation norm on bounded support.



When these two conditions are met, the targeted substitution estimator $\Psi(\hat P^*_n)$ exhibits asymptotic linearity with respect to the EIF $D(P)$. Moreover, by the Central Limit Theorem, $\Psi(\hat P^*_n)$ of each subgroup possesses asymptotic normality:
\begin{equation*}
\sqrt{n}\left\{\Psi(\hat P_n^*)-\Psi(P)\right\}\xrightarrow{\mathcal{D}} N\left[0,E_{P}\{D(P)^2\}\right].
\end{equation*}
The asymptotic variance of this estimator can be estimated by $\hat \sigma ^2=\frac{1}{n}\sum_{i=1}^n\{D(\hat P_n^*)\}^2$. The Wald confidence interval can be constructed by $\Psi(\hat P_n^*) \pm z_{1-\alpha/2}\sqrt{\mathbb{P}_n\{D(\hat P_n^*)\}^2}$, where $z_{1-\alpha/2}$ is the $(1-\alpha/2)^{th}$ percentile of the standard normal distribution. The result above is pointwise at a prespecified time point \(t_0\).  Under uniform versions of these regularity conditions, including \(\sup_{t\in[0,\tau]} |R_t(\widehat P_n^*,P)|=o_p(n^{-1/2})\), the asymptotic normality result could be extended to weak convergence of the time-indexed estimator, enabling simultaneous confidence bands over time. We leave this extension to future work.
 \vspace{-2pt}
\subsection{Variable importance}\label{c2ss:vi}
To identify key predictive variables driving treatment effect heterogeneity and to assess the prognostic variables most influential for treatment effect estimation within each subgroup, we compute two post-estimation variable importance measures (VIM). 

The first measure quantifies the \emph{predictive importance} of each variable in $\mathcal{V}$ by its contribution to the variance of the estimated individual-level treatment effects, following the idea of \citet{hines2022variable}. For the $j^{th}$ predictive variable ($j=1,2,\ldots,d$, $d < p$; recall $V \in \mathcal{V} \subset \mathcal{X} \subseteq \mathbb{R}^p$), 
\begin{equation}\label{vim1}
\begin{aligned}
    \text{VIM}_{1,j}&=\frac{\Bigr| \text{var}\left\{\hat \tau(V_i,L_i)\right\}-\text{var}\left\{\hat\tau(V_{i,-j},L_i)\right\}\Bigr|}{\text{var}\{\hat \tau(V_i,L_i)\}},
\end{aligned}
\end{equation} 
where $\hat\tau(V_{i},L_i)$ is the estimated treatment effect for individual $i$ using all predictive and prognostic variables, and $\hat \tau(V_{i,-j},L_i)$ is the estimate omitting the $j^{th}$ predictive variable. Larger $\text{VIM}_{1,j}$ values indicate greater contribution of the variable to treatment effect heterogeneity. 

The second measure quantifies the \emph{prognostic importance} of each variable in $\mathcal{L}$ by the absolute change in the estimated subgroup ATE when that variable is omitted, analogous to \citet{pirracchio2018collaborative}. For the $k^{th}$ prognostic variable ($k=1,2,\ldots,q$, $q \leq p-d$; recall $L \in \mathcal{L} \subseteq \mathcal{X} \setminus \mathcal{V}$),  
\begin{equation}\label{vim2}
    {\text{VIM}}_{2,k}=|\Psi(\hat P_n^*)-\Psi(\hat P^*_{n,-k})|,
\end{equation}
where $\Psi(\hat P_n^*)$ is the estimated subgroup ATE using all variables, and $\Psi(\hat P^*_{n,-k})$ is the corresponding estimate with the $k^{th}$ prognostic variable omitted ($k=1, \dots, q, q\leq (p-d)$. Larger $\text{VIM}_{2,k}$ values indicate greater impact of the variable on subgroup ATE estimation accuracy. 
Because the VIMs are defined at \(t_0\), variable-importance rankings may vary across follow-up times and should be interpreted for the chosen time horizon.
\vspace{-2pt}
\section{Simulation studies} \label{c2s:sim}
\subsection {Simulation setup}
We conducted simulation studies across various scenarios to evaluate the performance of our proposed TMLE framework. The R code is available on GitHub (link provided in the Supplementary Materials). We simulated two-cause competing risks data mimicking sepsis patients, based on a scheme similar to \citet{fine1999proportional}. The simulated data included binary predictive covariates $V_1$, $V_2$, and multiple prognostic covariates $L_1, L_2,\ldots, L_8$, with treatment assignment $A$ and competing event times $T_1$, $T_2$ generated from parametric models incorporating covariates and interaction terms (details in Web Appendix E).

To evaluate the double robustness and asymptotic properties of our proposed estimator, we designed scenarios with correct and misspecified models for outcome, treatment, and censoring. Each misspecified model deliberately omitted relevant covariates while introducing spurious ones (see Web Appendix E). Simulations were performed with sample sizes $N$ of ${800, 1500, 3000}$ and evaluated at the 25th, 50th, and 75th percentile event time.

To benchmark our estimator, we implemented S-learner and T-learner meta-algorithms \citep{kunzel_meta-learners_2019}. These were also used to generate initial estimates in our TMLE procedure. We compared the TMLE with a one-step estimator derived from the efficient influence function (EIF) (see Web Appendix F).

We also evaluated robustness to noise by adding 20 independent covariates, $L_9, L_{10},\ldots, L_{28}$, unrelated to outcome, treatment, or censoring, and examined performance when models were fit using either parametric or ensemble learners. Outcome models included standard or LASSO-regularized subdistribution hazards models (\texttt{fastcmprsk}), whereas treatment and censoring models were built using logistic regression, Cox models, or flexible Super Learner frameworks (\texttt{SuperLearner}, \texttt{survSuperLearner}) (details in Web Appendix E). This design highlights the flexibility of our TMLE framework in incorporating advanced machine learning methods alongside traditional models.

We estimated subgroup ATEs for each of the four groups defined by $V_1$ and $V_2$, and evaluated the performance using bias, RMSE, and empirical coverage of 95\% confidence intervals (details in Web Appendix E). We conducted $B = 500$ simulations for each scenario.

\subsection{Simulation results}
\begin{figure}
\centering
\includegraphics[width=0.84\textwidth]{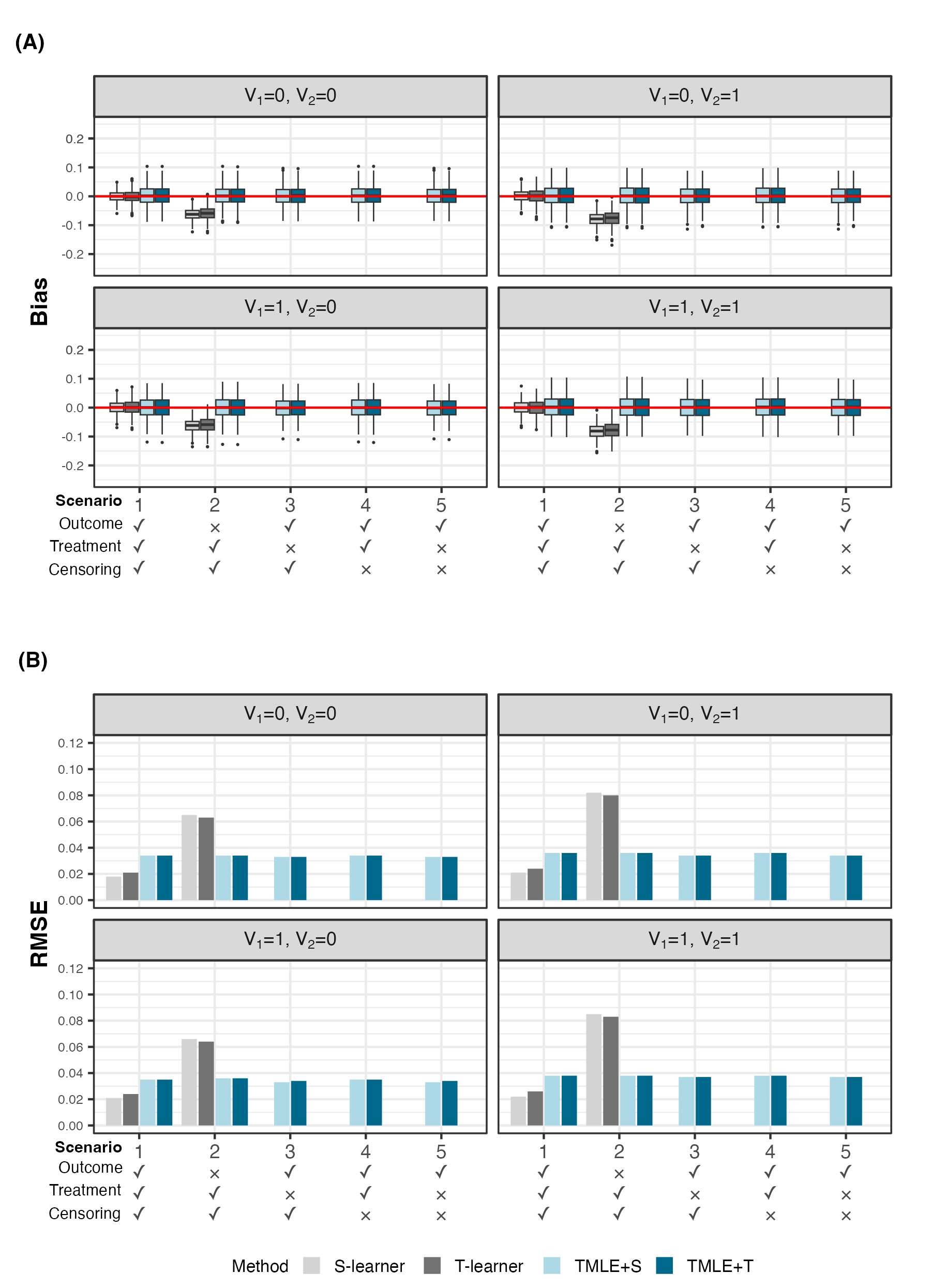}
\caption{\small Bias (upper panels) and RMSE (lower panels) of subgroup ATE estimations at the median follow-up time point under different model specifications. Each panel corresponds to one of the four subgroups determined by two binary predictive variables, $V_1$ and $V_2$. The red line represents zero bias. $\checkmark$ and $\times$ indicate that the corresponding models were correctly specified and misspecified, respectively. Light grey and dark grey represent S-learner and T-learner, respectively, while light blue and dark blue represent TMLE with S-learner and T-learner employed for the initial subdistribution hazard model in the first step, respectively. Treatment and censoring models are not involved in S-learner and T-learner. Sample size is $n=3000$ and each scenario was replicated 500 times.}\label{fig:BiasRMSE}
\end{figure}

Our simulation results confirm the asymptotic normality and double robustness of our subgroup ATE estimator. As shown in Figure \ref{fig:BiasRMSE}, our TMLE estimator outperforms S- and T-learners, especially when outcome or treatment models are misspecified, with lower bias and RMSE. In Scenario 1, where all models are correctly specified, all methods perform well with minimal bias and RMSE. However, in Scenario 2, with a misspecified outcome model, the S- and T-learners provide heavily biased estimates and exhibit a 200\% increase in RMSE compared to Scenario 1. In contrast, our TMLE framework produces unbiased estimates with the same RMSE as in Scenario 1. Similarly, in Scenarios 3-5, where the treatment and/or censoring models are misspecified, our TMLE methods show no inflation in bias and RMSE relative to Scenarios 1 and 2 (S- and T- learner do not involve treatment model so the results are not reported for Scenarios 3 and 4).We also evaluate performance under smaller samples ($n=1500, 800$) and at later time points (Web Appendix G). As expected, bias and RMSE increase as sample size decreases and time progresses, yet our methods remain more stable and accurate than alternatives.

Maintaining low bias and RMSE under misspecification of either the outcome or the treatment and censoring models demonstrates the double robustness of our method, confirming its value for estimating subgroup ATEs and guiding treatment decisions.


Table \ref{tab:coverage} reports the proportion of true subgroup ATEs covered by 95\% confidence intervals (CIs) from our TMLE method with the S-learner. Across all scenarios, with or without model misspecification, our method consistently achieves approximately 95\% coverage of the true subgroup ATE. These CIs, established by the asymptotic normality of our estimator, provide reliable inference for subgroup ATEs.

\begin{table}
\caption{Coverage probability of subgroup ATE estimations at the median follow-up time point using TMLE method, across various sample sizes and simulation scenarios$^a$.}\label{tab:coverage}
\small
\centering
\rowcolors{3}{gray!8}{white}
\begin{tabular}{c c c c c c}
\toprule
\text { Subgroup$^b$} & 
$\begin{array}{c}\textbf {Scenario 1$^c$} \\
\text {Outcome \checkmark} \\
\text {Treatment \checkmark}  \\
\text {Censoring \checkmark}
\end{array}$ &
$\begin{array}{c} \textbf {Scenario 2$^c$}  \\
\text {Outcome } \times\\
\text {Treatment } \checkmark \\
\text {Censoring } \checkmark
\end{array}$ & $
\begin{array}{c}\textbf {Scenario 3$^c$} \\
\text {Outcome } \checkmark \\
\text {Treatment }  \times\\
\text {Censoring \checkmark }
\end{array}$
& $\begin{array}{c}\textbf {Scenario 4$^c$} \\
\text {Outcome }  \checkmark\\
\text {Treatment }  \checkmark\\
\text {Censoring }  \times
\end{array}$ 
& $\begin{array}{c}\textbf {Scenario 5$^c$} \\
\text {Outcome }  \checkmark\\
\text {Treatment }  \times\\
\text {Censoring }  \times
\end{array}$ \\
\hline 
\rowcolor{gray!30}
\multicolumn{6}{c}{$n=3,000$} \\
$V_1=0, V_2=0$ & 95.4\% & 95.6\% & 96.0\% & 95.4\% & 96.0\% \\
$V_1=0, V_2=1$ & 96.2\% & 96.6\% & 97.2\% & 96.0\% & 97.2\% \\
$V_1=1, V_2=0$ & 94.6\% & 95.2\% & 94.4\% & 94.4\% & 94.4\% \\
$V_1=1, V_2=1$ & 96.6\% & 96.4\% & 96.0\% & 96.4\% & 96.0\% \\
\hline 
\rowcolor{gray!30}
\multicolumn{6}{c}{$n=1,500$} \\
$V_1=0, V_2=0$ & 95.6\% & 95.8\% & 95.6\% & 95.8\% & 95.6\% \\ 
$V_1=0, V_2=1$ & 96.2\% & 96.6\% & 98.0\% & 96.0\% & 98.0\% \\ 
$V_1=1, V_2=0$ & 93.8\% & 96.0\% & 94.0\% & 93.2\% & 94.0\% \\ 
$V_1=1, V_2=1$ & 95.4\% & 96.0\% & 96.0\% & 95.0\% & 96.0\% \\
\hline 
\rowcolor{gray!30}
\multicolumn{6}{c}{$n=800$} \\
$V_1=0, V_2=0$  & 93.0\% & 94.0\% & 94.6\% & 92.4\% & 94.6\% \\ 
$V_1=0, V_2=1$  & 96.6\% & 96.8\% & 98.6\% & 97.0\% & 98.6\%\\ 
$V_1=1, V_2=0$ & 93.6\% & 94.6\% & 94.8\% & 93.2\% & 94.8\% \\ 
$V_1=1, V_2=1$ & 95.2\% & 95.4\% & 95.8\% & 95.0\% & 95.8\% \\ 
\bottomrule
\end{tabular}
\\
\RaggedRight {$^a$$B=500$ Monte-Carlo repetitions were implemented. We reported results of TMLE method with S-learner for the initial outcome model, as other methods have similar coverage probability.\\
$^b$ Subgroups were defined by the two binary predictive variables $V_1$ and $V_2$. \\
$^c$$\checkmark$ indicates that the corresponding model was correctly specified, while $\times$ indicates that the model was misspecified.\\
\text{}}
\end{table}

\begin{figure}  
\centerline{\includegraphics[width=0.78\textwidth]{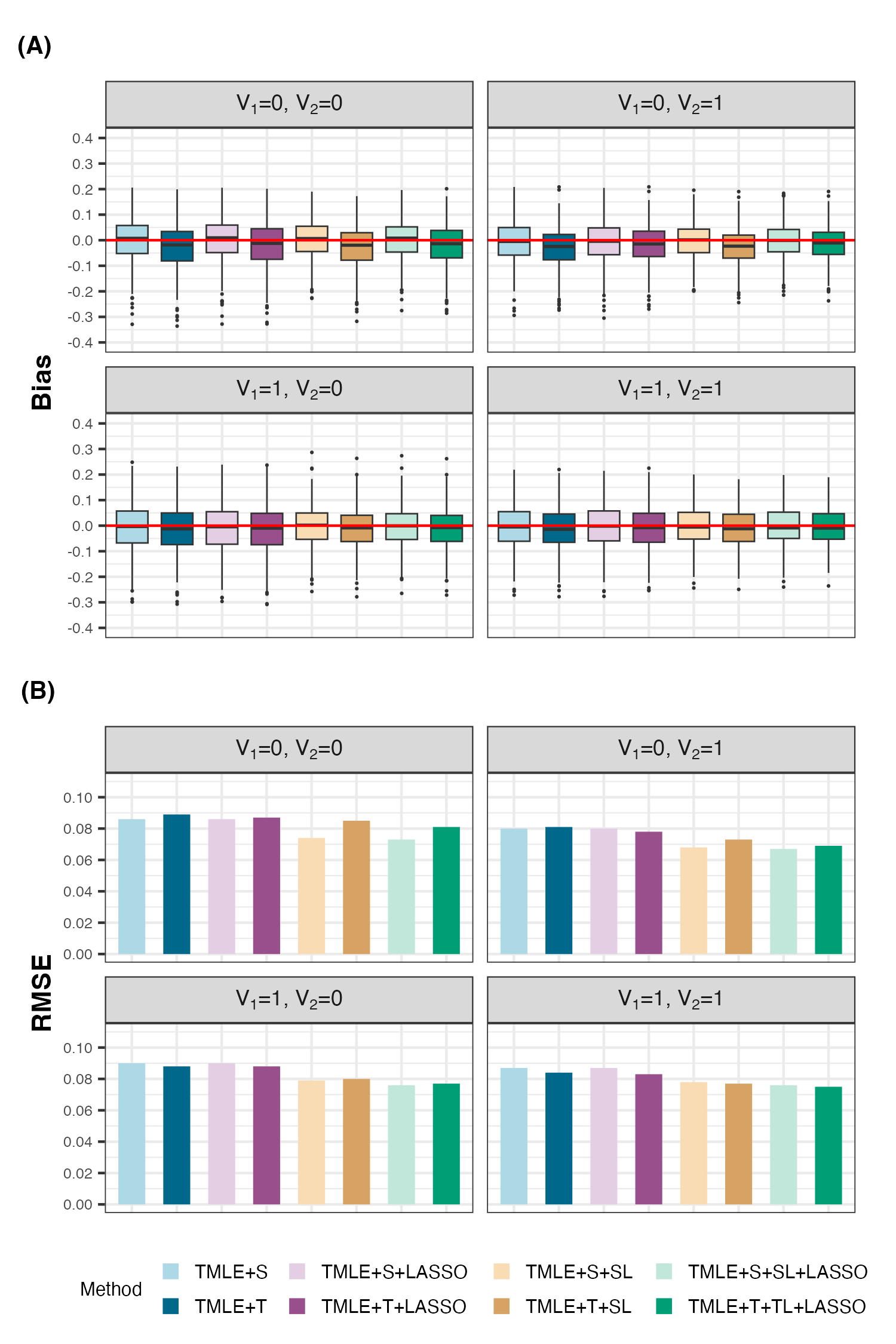}}
\caption[Performance of subgroup ATE estimations at the median follow-up time point in high-dimensional data.]{\small Bias (upper panels) and RMSE (lower panels) of subgroup ATE estimations at the median follow-up time point, using data with 30 covariates, 21 of which are noise variables. Each panel corresponds to one of the four subgroups determined by two binary predictive variables, $V_1$ and $V_2$. The red line represents zero bias. "+S" (light colors) and "+T" (dark colors) represent TMLE with S-learner and T-learner employed for the initial subdistribution hazard model in the first step, respectively; "+LASSO" represents using LASSO penalized subdistribution hazards model for the outcome; +SL" represents utilizing super learner for treatment and censoring. Sample size is $n=800$ and each scenario was replicated 500 times.}\label{fig:noises}
\end{figure} 
Figure \ref{fig:noises} shows the bias and RMSE of our proposed estimator in smaller samples (N=800) with irrelevant covariates, comparing different models for the outcome, treatment and censoring. In the upper panel, our proposed estimator provides unbiased estimates regardless of the model used. The number of outliers (upper panel) and the RMSE (lower panel) decrease when super learner are applied to treatment and censoring. Using a LASSO-regularized subdistribution hazards model for the CIF does not substantially reduce bias or RMSE.

Web Tables 1–4 report the performance of models for estimating subgroup ATEs across time points and sample sizes, compared with another doubly robust estimator—the one-step estimator derived from our EIF. In large samples ($N=3000$), the one-step estimator performs similarly to TMLE, but at smaller sizes, especially $N=800$, it shows higher RMSE. This is expected in small datasets where positivity is nearly violated and nuisance parameter estimates can become extreme, leading one-step estimators to produce values outside the expected range. Further details are provided in Web Appendix F.

\section{Analysis of Sepsis Dataset}\label{c2s:app}
\subsection{Data source and analysis}
We analyzed electronic medical records data from 2012 to 2018, including 3,245 ICU patients with sepsis who met ADRENAL trial eligibility criteria \citep{venkatesh2018adjunctive}. Our primary aim was to assess the impact of steroids on reducing ICU mortality among various sepsis patients. To focus on short-term treatment effects and mitigate potential violations of the positivity assumption, we considered follow-up was capped at 10 days post-ICU admission, with discharge by this time considered a competing event. 

Within the cohort, 885 patients (27.3\%) died within 10 ICU days, with 274 (31.0\%) receiving steroids. Meanwhile, 1,557 (48.0\%) were discharged within the first 10 days, including 188 (12.1\%) who received steroids. Treated patients had a median time to death of 1.83 days, compared to 2.79 days for untreated patients.

As presented in Web Table 5 in Supplementary Materials, we analyzed baseline characteristics of eligible patients, including demographics, comorbidities, clinical indices, and medication status. To identify potential predictive variables, a preliminary subdistribution hazards model was fitted with treatment and all covariates, revealing age ($<65$ vs. $\geq 65$), surgical history, and Glasgow Coma Scale (GCS, $<7$ vs. $7-15$)) as predictive variables with treatment effect heterogeneity (see cumulative incidence plots in Web Figure 1). Conditional on these three predictive variables, we considered gender, BMI, Sequential Organ Failure Assessment (SOFA) score, Charlson Comorbidity Index (CCI), and serum lactate as prognostic variables for estimating treatment effects within each subgroup. R code for data analysis is available on Github as the link provided in Supplementary Materials.

\subsection{Data analysis results}
We estimated subgroup ATE for each subgroup using our proposed TMLE framework and compared the subgroup ATEs estimated using the S-learner. For the TMLE, we reported 95\% CI based on the empirical EIF, while S-learner relied on bootstrapping. 

\begin{figure}
\centerline{\includegraphics[width=6in]{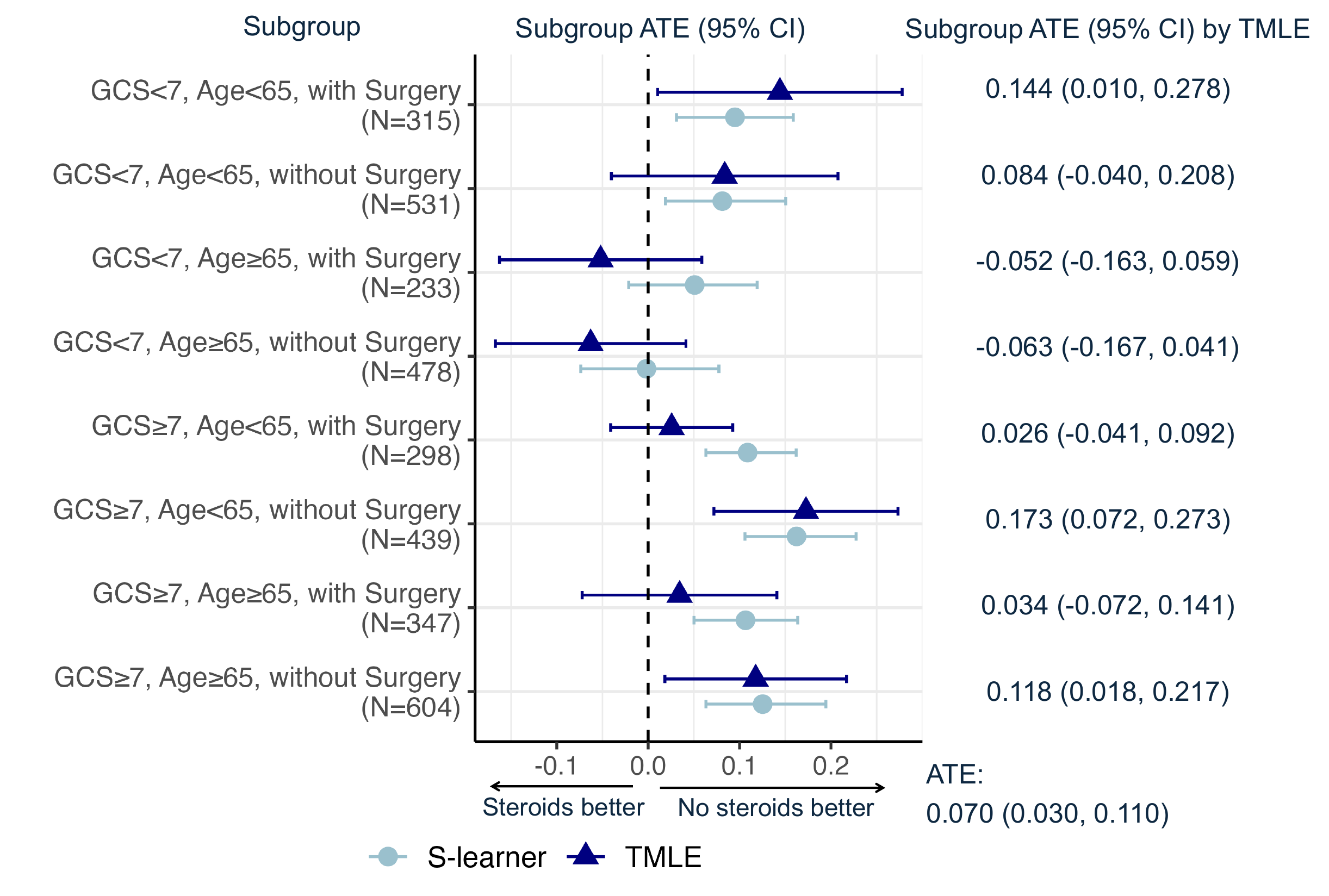}}
\caption{Subgroup average treatment effect (ATE) defined using the cumulative incidence function (CIF) at day 7 for patient subgroups categorized by age ($<$65 or $\ge$65 years old), surgical history (with or without surgery), and Glasgow Coma Scale (GCS). Subgroup ATE values were estimated by the proposed targeted maximum likelihood estimation (TMLE) framework (black rectangle) and the S-learner (grey circle). The 95\% confidence intervals are provided using the efficient influence function for TMLE, and through bootstrapping for the S-learner. The dashed line indicates no treatment effect.}\label{fig:sepsis_cate}
\end{figure}

Figure \ref{fig:sepsis_cate} summarizes treatment effect heterogeneity, providing insight about how the effect of steroids may differ across distinct patient subgroups. Positive estimates indicate increased risk of 7-day ICU mortality with steroid treatment, while negative values suggest a protective effect. Notably, patients who did not undergo surgery with GCS of 7 or higher (relatively higher consciousness level) showed estimated subgroup ATE significantly greater than 0, suggesting that the use of steroids does not benefit these patient subgroups. Among these patients, those aged 65 or older had an estimated subgroup ATE of 0.118 (95\% CI: 0.018-0.217), while younger patients had an estimate of 0.173 (95\% CI: 0.072-0.273). Similarly, for patients younger than 65 underwent surgery with GCS below 7, the estimated subgroup ATE was 0.144 (95\% CI: 0.010-0.278), also indicating a significant increase in the cumulative incidence of death on day 7 with steroid treatment compared to placebo.

However, for patients aged 65 or older with a GCS below 7 (relatively low consciousness level), regardless of surgery status, the estimated subgroup ATE were negative, with the 95\% CI including zero. Although there is some uncertainty, our method suggests the possibility of a beneficial effect of steroid treatment in these patient subgroups. Further investigations are needed to confirm potential benefits of steroids in this context.

Estimates from our TMLE framework differed from S-learner’s estimates in various groups, likely reflecting the absence of the doubly robust feature in the S-learner, and the uncertainty regarding correct model specification for outcome, treatment, and censoring. 

\begin{figure} \protect
\centerline{\includegraphics[width=5.5in]{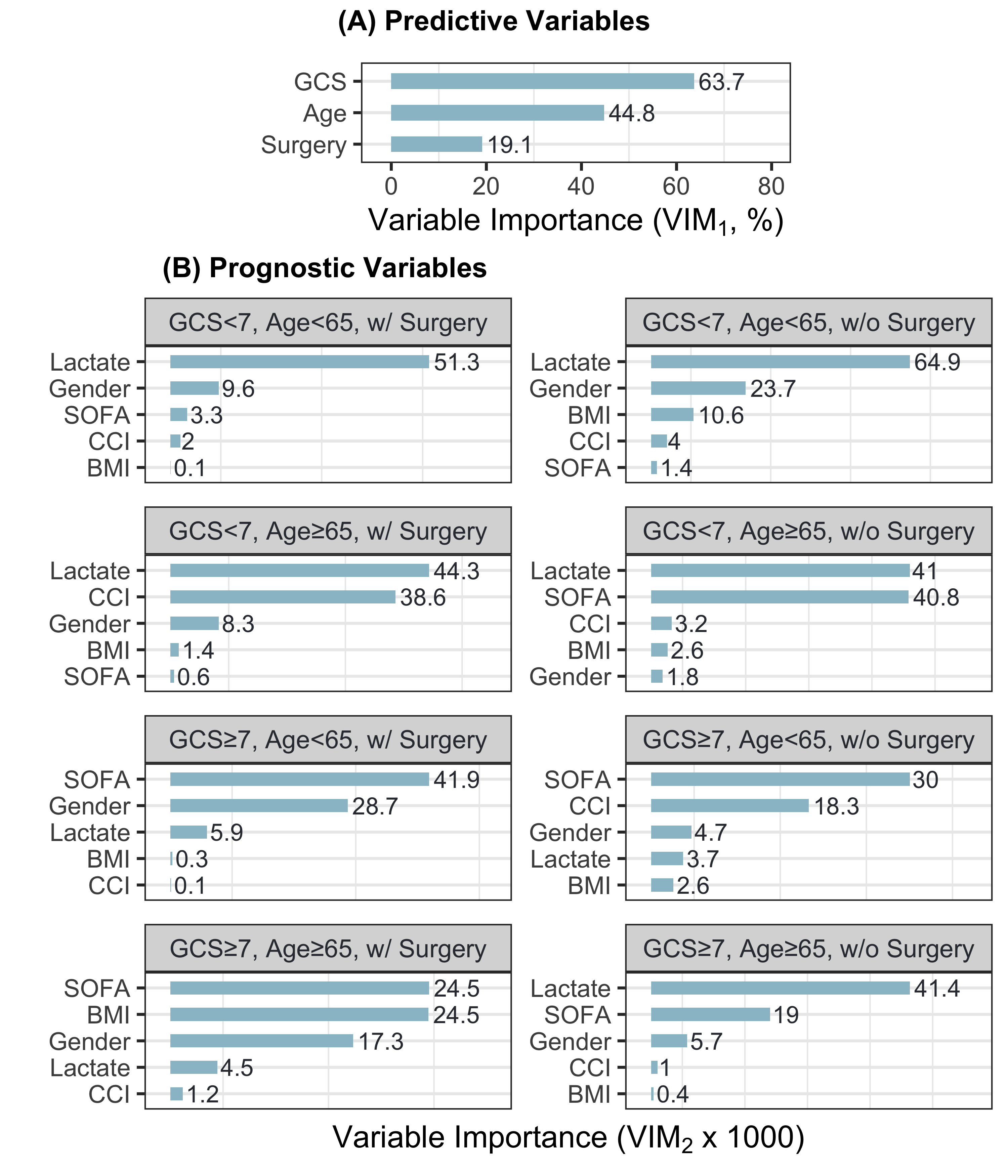}}
\caption{Variable importance for  predictive (A) and prognostic (B) variables. In (A), variable importance as estimated by $\text{VIM}_1$, represents the difference in empirical variance of the treatment effect when each variable is ignored during analysis. In (B), variable importance as estimated by $\text{VIM}_2$, shows the difference in the estimated treatment effect when each variable is removed from the analysis.}\label{fig:vim_all}
\end{figure}

To explore sources of treatment effect heterogeneity and identify variables relevant for subgroup ATE estimation, we calculated Variable Importance Measures (VIMs) for both predictive and prognostic variables within each subgroup. In Figure \ref{fig:vim_all}(A), we illustrate the variable importance of predictive variables, age, GCS and surgical history.  They are calculated using the difference in empirical treatment effect variance ($\text{VIM}_1$ from equation \ref{vim1}). Notably, GCS emerges as the most influential predictive variable, explaining the majority of variance in treatment effects. Omitting GCS from the analysis would reduce treatment effect variance by 63.7\%. Age was the next most important predictor, whereas surgical history contributed less. Prognostic variables had negligible influence on treatment effect heterogeneity compared with predictive variables (Web Figure 2, Supplementary Materials).

In Figure \ref{fig:vim_all}(B), we demonstrate the variable importance of prognostic variables, calculated using the differences in estimated treatment effects ($\text{VIM}_2$ in equation \ref{vim2}). For instance, in patients under 65 who underwent surgery and had a GCS $<7$, omitting serum lactate would result in a 0.0513 difference in treatment effect estimates, which is 36\% of the subgroup ATE (0.144). Serum lactate proves vital for estimating treatment effects in patients with lower consciousness levels (GCS $<7$), while SOFA is more important for patients with relatively higher consciousness levels (GCS $\geq 7$). This underscores the importance of collecting these variables for different groups to obtain accurate subgroup ATE estimates. 

\section{Discussion} \label{c2s:dis}
We developed a targeted maximum likelihood estimation (TMLE) framework for estimating subgroup average treatment effects (ATE) in time-to-event data with competing risks. Unlike approaches using cause-specific hazards as the fluctuation model, our framework directly targets the cumulative incidence function (CIF) by updating the subdistribution hazard. This choice aligns the estimation procedure with the primary prognosis-oriented estimand, enables direct interpretation of covariate effects on CIF, and improves computational efficiency.

Our estimator has a valuable doubly robust property, producing consistent estimates if either the outcome model or both the treatment and censoring models are correctly specified. These properties are established both theoretically and through extensive simulation studies across diverse scenarios.  It is a substitution estimator, and thus is guaranteed to be bounded within the interval (-1,1). Using our TMLE framework for treatment decisions enhances strategy reliability, even under model misspecification, for example, due to limited knowledge or data, which is very likely in most real-world problems.

We derived the efficient influence function (EIF) for the subgroup-specific CIF-based ATE and used it to construct valid statistical inference procedures, including confidence intervals and hypothesis tests. Our TMLE framework is highly flexible, allowing the use of various regression and machine learning algorithms for the outcomes, treatment, and censoring models, making it suitable for a wide range of data structures, including those with small sample sizes or many irrelevant covariates.

An important innovation of this work is the development of two complementary variable importance measures (VIMs): one quantifies the \textit{predictive importance} of each variable in explaining treatment effect heterogeneity, and the other measures the \textit{prognostic importance} of each variable in improving the precision of subgroup ATE estimation. These VIMs provide interpretable and clinically relevant tools for prioritizing variables in precision medicine. 

Future work will focus on integrating variable selection and subgroup ATE estimation into a unified procedure. In high-dimensional settings, where approaches such as LASSO may be inadequate, we will explore advanced machine learning methods (e.g., deep neural networks) to improve CIF estimation while preserving the double robustness.

\backmatter

\section*{Acknowledgements}
The sepsis data in this paper was supported by KNIGHTS R01 grant, grant number NIGMS 5R01GM141081.
\vspace{-1em}

\section*{Supporting Information}
Web Appendices, Tables, Figures, and R code referenced in Sections \ref{c2s:method}, \ref{c2sss:EIF}, \ref{c2sss:submodel}, \ref{c2ss:properties}, \ref{c2s:sim}, and \ref{c2s:app} are available with this paper at the \textit{Biometrics} website on Oxford Academic. The R code for the simulation study is available on GitHub (\text{https://github.com/Runjia-Li/SubgroupATE\_CR\_TMLE}).\vspace*{-8pt}

\section*{Data Availability}
Data sharing is not applicable as no new data were created or analyzed in this paper.
\bibliographystyle{biom} 
\bibliography{ref.bib}

\label{lastpage}

\end{document}


\label{firstpage}

\begin{abstract}
Supplementary material contains (A) the algorithm of TMLE for time-to-event outcome with competing risks, (B) the derivation of the efficient influence function, (C) the derivation of the score function of the fluctuation model, (D) a discussion regarding asymptotic properties based on the derivation of the second-order remainder, (E) a discussion on one-step estimator, (F) additional simulation results, and (G) additional analysis results of sepsis dataset. 

\end{abstract}
\maketitle

\section*{\textbf{Web Appendix A: Algorithm of TMLE for Time-to-Event Outcome with Competing Risks}}\label{app:algorithm}
\begin{algorithm}
\caption{}
\begin{algorithmic} 
\item \textbf{Preliminary step:} \\
\begin{enumerate} 
    \item Fit a subdistribution hazards model that includes all baseline covariates, treatment, and interaction terms between treatment and covariates. 
    \item Identify predictive covariates based on the fitted regression results and clinical expertise. 
    \item Split data into $M$ subgroups based the predictive variables.
\end{enumerate} \vspace{0.5em}
\item \textbf{Within each subgroup $m=1,2,...,M$:} \vspace{0.3em}
\item [\textbf{1.}] \textbf{Initial estimation step}  
\begin{enumerate} 
    \item Subdistribution hazard $\hat \lambda_{1,m}^{(0)}(t|A,L)$ using unpenalized or penalized subdistribution hazards model, and CIF as $\hat{F}_{1, m}^{(0)}\left(t | A, L\right)=1-\exp\{-\int^{t}_0 \hat\lambda^{(0)}_{1,m}(t|A,L)dt\}$;
    \item Treatment assignment $\hat \pi_m(t|A,L)$ by logistic regression, or super learner;
    \item Censoring probability $\hat G_m(t|A,L)$ by Cox regression, or super learner.
\end{enumerate}
\item [\textbf{2.}]\textbf{Targeting step} \\
Update estimated subdistribution hazard and CIF. Let $k=1,2, \ldots$ denote the iteration index. \\ \quad \quad Initiate $\hat \epsilon=1$. Let $s_n$ be a small number, e.g., 0.001.\\
\quad \quad \textbf{while} {$|\hat \epsilon| > s_n$} \textbf{do} \\
\quad \quad \quad \textbf{(1)} Obtain $\hat \epsilon$ by maximizing the partial likelihood with respect to the model:
$$\lambda_{1, m, \epsilon}^{(k)}\left(t | A, L\right) =\lambda_{1, m}^{(k)}\left(t | A, L\right) \exp \left\{\epsilon h_{t_0,m}^{(k)}\left(t,A, L\right)\right\},$$
 \quad \quad \quad \quad \quad where the clever covariate $h_{m}^{(k)}(t,A,L)=\frac{\{\mathbb{I}(A=1)-\mathbb{I}(A=0)\}}{\hat \pi_m(A | L)} \cdot \frac{\{1-\hat F_{1,m}^{(k)}\left(t_{0} | A, L\right)\}}{\{1-\hat F_{1,m} ^{(k)}(t | A, L)\} \hat G_m(t | A, L)}.$\\
 \quad \quad \quad\textbf{(2)} Update estimators:
$$\hat{\lambda}_{1, m}^{(k+1)}\left(t | A, L\right)=\hat{\lambda}_{1, m}^{(k)}\left(t | A, L\right) \exp \left\{\hat{\epsilon} h_{m}^{(k)}\left(t, A, L\right)\right\}\text{, and}$$  
$$\hat{F}_{1, m}^{(k+1)}\left(t | A, L\right)=1-\exp \left\{-\int_0^{t} \hat{\lambda}_{1, m}^{(k+1)}\left(t | A, L\right) d t\right\}.$$
\quad  \quad \quad \quad  $k=k+1$
\item \quad \quad \textbf{end while}
\item [\textbf{3.}]\textbf{Subgroup ATE estimation:}
\item \quad \quad Denote the final estimators obtained from Step 2 as $\hat F_{1,m}^*(t|A,L)$. Plug in the estimates and calculate the average treatment effect within each subgroup $m$ to obtain the subgroup ATE:
\quad\quad $$\Psi_{t_0,m}(\hat P_n^*)=\frac{1}{n_m}\sum_{i=1}^{n_m}\left\{(\hat F_{1,m}^*(t_0|L_i,A_i=1)-\hat F_{1,m}^*(t_0|L_i,A_i=0)\right\}.$$\vspace{-0.5em}
\end{algorithmic}\label{c2a:tmle}
\end{algorithm}

\normalsize
\newpage
\section*{\textbf{Web Appendix B: Derivation of the Efficient Influence Function}}\label{app:c2thm:eif}
In this section and the following sections, we drop the subscript $m$ and $t_0$ for notational simplicity, with the understanding that the entire procedure is carried out separately for each subgroup. Our procedure targets estimation and pointwise inference within each subgroup $V\in \mathcal{V}_m$, rather than uniform inference across all subgroups. 

The EIF is derived by calculating the pathwise derivative $\frac{d}{d\epsilon} \Psi(P_\epsilon)\Bigr|_{\epsilon=0}$ for a submodel $P_\epsilon$: $dP_\epsilon=\{1+\epsilon b\}dP$. Our goal is to express the pathwise derivative of the target estimand as an inner product of two terms, i.e.,
 $\frac{d}{d\epsilon} \Psi(P_\epsilon)\Bigr|_{\epsilon=0}=E\{D(P)\cdot b\}$.

Under Assumptions 1-4, our target estimand, subgroup ATE, can be identified and expressed as
$$
\begin{aligned}
\Psi(P)
&=\int_{\mathcal{L}}\sum_{a=0,1}F_{1}\left(t_{0}|  A=a,L=l\right)\left\{\mathbbm{1}(a=1)-\mathbbm{1}(a=0)\right\}\mu(l)dl  \\
& =\int_{\mathcal{L}} \sum_{a=0,1}\left\{1-\exp \left(-\int_{0}^{t_{0}} \Lambda_{1}(d t |  a,l)\right)\right\} \left\{\mathbbm{1}(a=1)-\mathbbm{1}(a=0)\right\} \mu(l) d l,\\
\end{aligned}
$$
where $\mu(l)$ is the density of prognostic variables $L$, and
$$
\begin{aligned}
\quad \quad \quad \Lambda_{1}(d t |  a, l)
&= \frac{\operatorname{Pr}\{t\leq{T} < t+dt, {\Delta}=1 ,C\geq t|  a,l\}}{\operatorname{Pr}[\{{T}\geq t \cup (T\leq t \cap{\Delta} \neq 1)\}\cap C\geq t|   a,l] } \\
& =\frac{\operatorname{Pr}\{t\leq{T}<t+d t, {\Delta}=1 |  a, l\}\operatorname{Pr}\{C\geq t|  a,l\}}{\operatorname{Pr}\{{T} \geq t	\cup ({T} \leq t \cap \Delta \neq 1) |  a, l\}\operatorname{Pr}\{C\geq t|  a,l\}} \\
& =\frac{P^{1}(d t |  a, l)}{\sum _\delta\{\int_0^\infty P^{\delta}(d s |  a, l)-\int_{0}^{t}\mathbbm{1}(\delta=1)P^{\delta}(d s |  a, l)\}}, 
\end{aligned}
$$
where $P^\delta(d t | a,l)=\operatorname{Pr}(t\leq{T}< t+d t,  {\Delta}=\delta | a,l), \delta \in \{1,2\}$, assuming two causes for simplicity. 

The pathwise derivative of the target estimand
$$
 \begin{aligned}
&\frac{d}{d \epsilon}\Psi (P_{\epsilon}) \Bigr| _{\epsilon=0}  \\
& =\frac{d}{d \epsilon}\int_{\mathcal{L}} \sum_{a=0,1}\left\{1-\exp \left(-\int_{0}^{t_{0}} \Lambda_{1,\epsilon}(d t |  a,l)\right)\right\} \left\{\mathbbm{1}(a=1)-\mathbbm{1}(a=0)\right\} \mu_\epsilon(l) d l\Bigr| _{\epsilon=0}\\
& =\frac{d}{d \epsilon} \int_{\mathcal{L}} \sum_{a=0,1}\left[1-\exp \left\{-\int_{0}^{t_{0}} \Lambda_{1,\epsilon}(d t |  a,l)\right\}\right] \left\{\mathbbm{1}(a=1)-\mathbbm{1}(a=0)\right\} \mu_\epsilon(l) d l\Bigr| _{\epsilon=0} \\
& =\int_{\mathcal{L}} \sum_{a=0,1} \exp\left\{-\int_{0}^{t_{0}} \Lambda_{1}(d t |  a, l)\right\}\left\{\int_{0}^{t_{0}} \frac{d}{d \epsilon} \Lambda_{1,\epsilon}(d t |  a, l)\Bigr| _{\epsilon=0}\right\}\left\{\mathbbm{1}(a=1)-\mathbbm{1}(a=0)\right\}\mu(l)dl\\
& \quad\quad+ \int_{\mathcal{L}} \sum_{a=0,1} \left[1-\exp \left\{-\int_{0}^{t_{0}} \Lambda_1(d t |  a, l)\right\}\right] \left\{\mathbbm{1}(a=1)-\mathbbm{1}(a=0)\right\}\frac{d}{d \epsilon} \mu_{\epsilon=0}\Bigr| _{\epsilon=0}dl\\
&:=(*),
\end{aligned}
$$

where
$$
\begin{aligned}
 &\frac{d}{d \epsilon} \Lambda_{1, \epsilon}(d t |  a, l)\Bigr| _{\epsilon=0} =\frac{d}{d \epsilon} \frac{P_\epsilon^{1}(d t |  a, l)}{P_\epsilon \{{T} \geq t	\cup ({T} \leq t \cap \Delta\neq 1) |  a, l\}} \Bigr| _{\epsilon=0}\\
&=P\{{T} \geq t \cup({T} \leq t \cap \Delta \neq 1 ) |  a, l\}^{-2} \cdot \\
&\quad\quad\left[P^{1}(d t |  a, l) b(t, 1, a, l) P\{{T} \geq t \cup({T} \leq t \cap  \Delta \neq 1) |  a, l\}\right.\\
& \quad\quad -P^{1}(d t |  a, l) \sum_{\delta=1,2} \int_0^\infty \{P^{\delta}(d s |  a, l)-\mathbbm{1}(T<t,\delta=1)\left.P^{\delta}(d s |  a, l)\} b(s, \delta, a, l)\right] \\
& =\sum_{\delta=1,2}\left[\frac{ \mathbbm{1}(\delta=1)P^{\delta}(d t |  a, l) b(t, \delta, a, l)}{P\{{T} \geq t \cup({T} \leq t \cap \Delta \neq 1 ) |  a, l\}} \right.\\
&\quad\quad-\left.\frac{\Lambda_1(d t |  a, l)  }{P\left\{{T} \geq t \cup({T} \leq t \cap \Delta \neq 1 ) |  a, l\right\}}\int_0^\infty Y(t)P^{\delta}(d s |  a, l)  b(s, \delta, a, l)\right]\text{, and}
\end{aligned}
$$
{\text{\quad by the key identity (1) in \cite{levy2019tutorial}}},
$$
\frac{d}{d \epsilon}\mu_\epsilon(l)\Bigr| _{\epsilon=0} 
=\left[E_P\left\{b(T,\Delta,A,L)| l\right\}-E_P\left\{b(T,\Delta,A,L)\right\}\right]\mu(l).
$$


$$
\begin{aligned}
\text{Let (1)}&= \int_{0}^{t_{0}}\frac{d}{d\epsilon} \Lambda_{1,\epsilon}(d t |  a, l)\Bigr|_{\epsilon=0}\\
&= \int_0^\infty\sum_{\delta=1,2} \mathbbm{1}(t\leq t_0)\left[\frac{\mathbbm{1}(\delta=1)P^{\delta}(d t |  a, l) b(t, \delta, a, l)}{P\left\{{T} \geq t \cup({T} \leq t \cap \Delta \neq 1 ) |  a, l\right\}}\right. \\
& \quad\quad-\left.\frac{\Lambda_1(d t |  a, l)  }{P\{{T} \geq t \cup({T} \leq t \cap \Delta \neq 1 ) |  a, l\}}\int_0^\infty Y(t)  P^{\delta}(d s |  a, l) b(s, \delta, a, l)\right] \\
&=\int_0^\infty\sum_{\delta=1,2} \left[\frac{\mathbbm{1}(t\leq t_0,\delta=1)}{P\{{T} \geq t \cup({T} \leq t \cap \Delta \neq 1 ) |  a, l\}}P^{\delta}(d t |  a, l) b(t, \delta, a, l)\right.\\
&\quad \quad -\int_0^\infty\left.\frac{\Lambda_1(d t |  a, l) \mathbbm{1}(t\leq t_0) }{P\{{T} \geq t \cup({T} \leq t \cap \Delta \neq 1 ) |  a, l\}} Y(t)  P^{\delta}(d s |  a, l) b(s, \delta, a, l)\right] \\
&=\int_0^\infty\sum_{\delta=1,2} \left[\int^{t_0}_0 \frac{N(dt)}{P\{{T} \geq t \cup({T} \leq t \cap \Delta \neq 1) |  a, l\}}P^{\delta}(d s |  a, l) b(s, \delta, a, l)\right.\\
&\quad \quad -\left.\int_0^{t_0}\frac{\Lambda_1(d t |  a, l) Y(t) }{P\{{T} \geq t \cup({T} \leq t \cap \Delta \neq 1) |  a, l\}}   P^{\delta}(d s |  a, l) b(s, \delta, a, l)\right] \\
&=\int_0^\infty\sum_{\delta=1,2} \left\{\int^{t_0}_0 \frac{N(dt)-\Lambda_1(d t |  a, l) Y(t)}{1-F_1(t| a,l)}\right\}P^{\delta}(d s |  a, l) b(s, \delta, a, l); \text{ and}
\end{aligned}
$$

$$
\begin{aligned}
\text{(2)}&= \int_{\mathcal{L}} \sum_{a=0,1} \left[1-\exp \left\{-\int_{0}^{t_{0}} \Lambda_1(d t |  a, l)\right\}\right] \left\{\mathbbm{1}(a=1)-\mathbbm{1}(a=0)\right\}\frac{d}{d \epsilon} \mu_{\epsilon=0}\Bigr| _{\epsilon=0}\mu(l) d l\\
&=\int_{\mathcal{L}} \sum_{a=0,1} \left[1-\exp \left\{-\int_{0}^{t_{0}} \Lambda_1(d t |  a, l)\right\}\right] \{\mathbbm{1}(a=1)-\mathbbm{1}(a=0)\}\cdot\\ &\quad\quad\left[E_P\left\{b(T,\Delta,A,L)| l\right\}-E_P\left\{b(T,\Delta,A,L)\right\}\right]\mu(l) d l\\
&=\int_{\mathcal{L}} \left[\left\{ 1-\exp \left(\int_{0}^{t_{0}} \Lambda_{1} (d t |  1, l) \right) \right\}\right.\\ &\quad\quad -\left.\left\{1-\exp \left(\int_{0}^{t_{0}}\Lambda_{1}(d t |  0, l)\right) \right\}\right] E_P\{b(T,\Delta,A,L)| l\}\mu(l) d l\\
&\quad \quad -\int_{\mathcal{L}} \left[\left\{ 1-\exp \left(\int_{0}^{t_{0}} \Lambda_{1} (d t |  1, l) \right) \right\}\right.\\ &\quad\quad -\left.\left\{1-\exp \left(\int_{0}^{t_{0}}\Lambda_{1}(d t |  0, l)\right) \right\}\right] E_P\{b(T,\Delta,A,L)\}\mu(l) d l\\
& = E_{P}\left[ \{F_{1}(t_{0} |  1,L)-F_{1}(t_{0} |   0,L) \}b({T},  \Delta, A, L) \right]-E_P\{\Psi(P)b({T},  \Delta, A, L) \}.
\end{aligned}
$$
\\
Plug (1) and (2) back into (*), we have
$$
\begin{aligned}
\frac{d}{d \epsilon}\Psi (P_{\epsilon}) \Bigr| _{\epsilon=0} =& \int_{\mathcal{L}} \sum_{a=0,1} \exp\left\{-\int_{0}^{t_{0}} \Lambda_{1}(d t |  a, l)\right\}\left\{\mathbbm{1}(a=1)-\mathbbm{1}(a=0)\right\}\\
&\quad \quad \left[\int_0^\infty\sum_{\delta=1,2} \left\{\int^{t_0}_0 \frac{N(dt)-\Lambda_1(d t |  a, l) Y(t)}{1-F_1(t| a,l)}\right\}P^{\delta}(d s |  a, l) b(s, \delta, a, l)\right]\mu(l)dl\\
& \quad\quad+ E_{P}\left[ \{F_{1}(t_{0} |  1,L)-F_{1}(t_{0} |   0,L) \}b({T},  \Delta, A, L) \right]-E_P\{\Psi(P)b({T},  \Delta, A, L) \}\\
=& \int_{\mathcal{L}} \sum_{a=0,1} \left\{1-F_{1}(t_0|  a, l)\right\}\left\{\frac{\mathbbm{1}(a=1)}{\pi(a| l)}-\frac{\mathbbm{1}(a=0)}{\pi(a| l)}\right\}\pi(a| l)\\
&\quad \quad \left[\int_0^\infty\sum_{\delta=1,2} \left\{\int^{t_0}_0 \frac{N(dt)-\Lambda_1(d t |  a, l) Y(t)}{1-F_1(t| a,l)}\right\}P^{\delta}(d s |  a, l) b(s, \delta, a, l)\right]\mu(l)dl\\
& \quad\quad+ E_{P}\left[ \{F_{1}(t_{0} |  1,L)-F_{1}(t_{0} |   0,L) \}b({T},  \Delta, A, L) \right]-E_P\{\Psi(P)b({T},  \Delta, A, L) \}\\
=& \int_{\mathcal{L}} \sum_{a=0,1}\pi(a| l)\int_0^\infty\sum_{\delta=1,2} \left\{\int^{t_0}_0\left[ \frac{\mathbbm{1}(a=1)-\mathbbm{1}(a=0)}{\pi(a| l)}\cdot\frac{ 1-F_{1}(t_0|  a, l)}{1-F_1(t| a,l)}\cdot\right.\right.\\
&\quad\quad\left.\left.\left\{N(dt)-\Lambda_1(d t |  a, l) Y(t)\right\}\right] P^{\delta}(d s |  a, l) b(s, \delta, a, l)\right\}\mu(l)dl\\
& \quad\quad+ E_{P}\left[ \{F_{1}(t_{0} |  1,L)-F_{1}(t_{0} |   0,L) \}b({T},  \Delta, A, L) \right]-E_P\{\Psi(P)b({T},  \Delta, A, L)\}
\end{aligned}
$$

$$
\begin{aligned}
=& E_P\left( \int^{t_0}_0 \left[\frac{\mathbbm{1}(A=1)-\mathbbm{1}(A=0)}{\pi(A| L)}\cdot\frac{ 1-F_{1}(t_0|  A, L)}{1-F_1(t| A,L)}\left\{N(dt)-\Lambda_1(d t |  A, L) Y(t)\right\}\right] b(T, \Delta, A, L)\right)\\
& \quad\quad+ E_{P}\left[ \left\{F_{1}(t_{0} |  1,L)-F_{1}(t_{0} |   0,L) \right\}b({T},  \Delta, A, L) \right]-E_P\left\{\Psi(P)b({T},  \Delta, A, L) \right\}\\
=& E_P\left\{\left[\int^{t_0}_0 \frac{\mathbbm{1}(A=1)-\mathbbm{1}(A=0)}{\pi(A| L)}\cdot\frac{ 1-F_{1}(t_0|  A, L)}{1-F_1(t| A,L)}\left\{N(dt)-\Lambda_1(d t |  A, L) Y(t)\right\}\right.\right.\\
& \quad\quad+\left.\left.\{F_{1}(t_{0} |  1,L)-F_{1}(t_{0} |   0,L) \}-\Psi(P)\right]b({T},  \Delta, A, L)\right\}. \\
\end{aligned}
$$
\\
Now we have shown that the pathwise derivative of the targeted estimand can be expressed as an inner product of $D(P)$ and $b({T},  \Delta, A, L) $.
Therefore, the EIF of $\Psi(P)$ for complete data (no censoring) is
$$
\begin{aligned}
D(P)&= \int^{t_0}_0 \frac{\mathbbm{1}(A=1)-\mathbbm{1}(A=0)}{\pi(A| L)}\cdot\frac{ 1-F_{1}(t_0|  A, L)}{1-F_1(t| A,L)}\left\{N(dt)-\Lambda_1(d t |  A, L) Y(t)\right\}\\
& \quad\quad+\{F_{1}(t_{0} |  1,L)-F_{1}(t_{0} |   0,L) \}-\Psi(P),
\end{aligned}
$$
with the clever covariate is
$$h(t,A,L)=\frac{\mathbbm{1}(A=1)-\mathbbm{1}(A=0)}{\pi(A| L)}\cdot\frac{ 1-F_{1}(t_0|  A, L)}{1-F_1(t| A,L)}.$$
For right-censored data, $N(t)$ and $\left\{1-Y(t)\right\}$ are not observable when $r(t)=\mathbbm{1}(C\geq T\wedge t)=0$, but $r(t)N(t)$ and $r(t)\left\{1-Y(t)\right\}$ are computable for $r(t)=0,1$ \citep{fine1999proportional}. The term $\frac{r(t)}{G\{(\tilde T\wedge t)^-| A,L\}}$ has an expectation 1 conditional of $T, \Delta,$ and $L$. We multiply the first part of EIF by $\frac{w(t)}{G(t^-|A,L)}$, then the EIF of $\Psi(P)$ for the right-censored data has the form:
$$
\begin{aligned}
D(P)&=\int^{t_0}_0 \frac{\mathbbm{1}(A=1)-\mathbbm{1}(A=0)}{\pi(A| L)}\cdot\frac{ 1-F_{1}(t_0|  A, L)}{1-F_1(t| A,L)}\cdot\frac{w(t)}{G(t^-|A,L)}\left\{N(dt)-\Lambda_1(d t |  A, L) Y(t)\right\}\\
& \quad\quad+\{F_{1}(t_{0} |  1,L)-F_{1}(t_{0} |   0,L) \}-\Psi(P),
\end{aligned}
$$
where $w(t)=\frac{\mathbbm{1}(C\geq T\wedge t)G(t^-|A,L)}{G\{(\tilde{T}\wedge t)^-| A,L\}}$ and the clever covariate being
$$h(t,A,L)=\frac{\mathbbm{1}(A=1)-\mathbbm{1}(A=0)}{\pi(A| L)G(t^-|A,L)}\cdot\frac{ 1-F_{1}(t_0|  A, L)}{1-F_1(t| A,L)}.$$

\newpage
\section*{\textbf{Web Appendix C: Derivation of the Score Function of the Fluctuation Model}}\label{app:c2lik}
\vspace{-1em}
The fluctuation model that we use for updating the estimated target parameter is
    $$\lambda_{1,\epsilon}(t| A,L)=\lambda_{1}(t|  A,L)\exp\{\epsilon h(t,A,L)\}.$$
The weighted likelihood (the pseudo likelihood (page 5, section 2.3) in \cite{bellach2019weighted}) has the form:
$$\mathscr{L}_n(\lambda_{1,\epsilon})=\prod_{i=1}^n \left[\left\{ w_i( t)Y_i(t)\lambda_{1,\epsilon}(t|  A_i,L_i)^{\mathbbm{1}(C_i\geq t) N_i(dt)} \right\}\exp \left\{-\int_{0}^{t_0} w_i(t) Y_i(t)\lambda_{1,\epsilon}(t| A_i,L_i)dt \right\}\right];$$
and the weighted log-likelihood function (page 8, section 3.2) in \cite{bellach2019weighted})
$$ 
\begin{aligned}
\log\mathscr{L}_n(\lambda_{1,\epsilon})&=\sum_{i=1}^n\left[\int_{0}^{t_0}\mathbbm{1}(C_i\geq t)Y_i(t)w_i(t)\log\left\{\lambda_{1,\epsilon}(t|A_i, L_i) \right\}N_i(dt)-\int_{0}^{t_0} w_i(t)Y_i(t)\lambda_{1,\epsilon}\left(t|A_i, L_i\right)dt\right].
\end{aligned}
$$
Since $N_i(dt)=N_i(t)-N_i(t-)=0$ as long as $\mathbbm{1}(C_i\geq t) Y_i(t)=0$, we can ignore $\mathbbm{1}(C_i\geq t) Y_i(t)$ in the first part of the likelihood. Therefore, the score function has the form:
$$ \begin{aligned}
\frac{d}{d \epsilon}\log\mathscr{L}_n\left(\lambda_{1, \epsilon}\right) &=\sum_{i=1}^n \left[\int_{0}^{t_0}w_i(t) \frac{d}{d \epsilon}\log \left\{\lambda_{1,\epsilon}(t |  A_i, L_i)\right\} N_i(dt) -\int_{0}^{t_0} w_i(t)Y_i(t) \frac{d}{d\epsilon} \lambda_{1,\epsilon}(t|A_i,L_i) dt \right] \\
&=\sum_{i=1}^n \left[\int_{0}^{t_0} w_i(t)\frac{\lambda_1(t|A_i,L_i)\exp\{\epsilon h(t,A_i,L_i)\}}{\lambda_1(t|A_i,L_i)\exp\{\epsilon h(t,A_i,L_i)\} }h(t,A_i,L_i) N_i(d t)\right.\\
&\quad \quad- \left.\int_{0}^{t_0} w_i(t)Y_i(t) \lambda_{1}(t|A_i,L_i)\exp\{\epsilon h(t,A_i,L_i)\}  h(t,A_i,L_i) d t \right]\\
&=\sum_{i=1}^n \int_{0}^{t_0} w_i(t)h(t,A_i,L_i) \left[N_i(d t)-Y_i(t) \lambda_{1}(t|A_i,L_i)\exp\{\epsilon h(t,A_i,L_i)\} d t\right].\\
\end{aligned}$$
Then,
$$\begin{aligned}
\frac{1}{n
}\cdot\frac{d}{d \epsilon}\log\mathscr{L}_n&\left(\lambda_{1, \epsilon}\right) \Bigr| _{\epsilon=0}=\frac{1}{n}\sum_{i=1}^n \int_{0}^{t_0} w_i(t)h(t,A_i,L_i) \left[N_i(d t)-Y_i(t) \lambda_{1}(t|A_i,L_i) d t\right]
\end{aligned}$$
is exactly the same as the empirical average of the first part of EIF. \\
Empirically, the maximum likelihood estimator $\hat \epsilon$ is obtained by using R package \texttt{nleqslv}.
\newpage
\section*{\textbf{Web Appendix D: Asymptotic Properties: Derivation of the Second-order Remainder}}\label{app:c2rem}
By the von Mises expansion (distributional Taylor expansion), 
\begin{equation}\label{eq:vonmis_sATE}
\begin{aligned}
\Psi(\hat P^*_n)-\Psi(P)
&=-E(D(\hat P^*_n))+R(\hat P^*_n,P) \\
&=\left[\mathbb{P}_n\left\{D(\hat P^*_n)-D(P)\right\}-E\left\{D(\hat P^*_n)-D(P)\right\}\right]\\
&\quad-\mathbb{P}_n(D(\hat P^*_n))+\mathbb{P}_n(D(P))\\
&\quad+R(\hat P^*_n,P),
\end{aligned}
\end{equation}
The first term is $o_p(n^{-1/2})$ if $D(\hat P^*_n)$ belongs to the Donsker class by the empirical process theory \citep{van2000asymptotic, van2011targeted}. The second term, the mean of EIF, is $o_p(n^{-1/2})$ by solving the efficient influence function equation in the targeting step (see (7)). The third term is $o_p(n^{-1/2})$ by the Central Limit Theorem. To show the asymptotic properties of the targeted estimator $\Psi(\hat P_n^*)$, we need to show that the last term, the second order remainder $R(\hat P_n^*,P_0)=o_P(n^{-1/2})$.\\
\\
The second-order remainder $R(\hat P^*_n, P)=\Psi(\hat P^*_n)-\Psi(P)+E\left\{D(\hat P^*_n)\right\},$\\
where
$$
\begin{aligned}
&E\left\{D(\hat P^*_n)\right\}\\
&=E\left[\int_0^{t_0}\hat h(t,A,L)\hat w(t) \left\{N(dt)-\hat\Lambda_1^*(d t|A, L) Y(t)\right\}+ \left\{\hat F_{1}^*(t_{0} | 1,L)-
\hat F^*_{1}(t_{0} | 0,L)\right\}-\Psi(\hat P)\right]\\
&=E\left[\int_0^{t_0} \frac{\{\mathbbm{1}(A=1)-\mathbbm{1}(A=0)\}}{\hat \pi(A|L)} \cdot \frac{\{1-\hat F^*_{1}\left(t_{0}|A,L\right)\}}{\{1-\hat F^*_1 (t|A,L)\}\hat G(t^-|A,L)}\right.\cdot\\
&\quad\quad\left.\frac{\mathbbm{1}(C\geq T\wedge t)\hat G(t^-|A,L)}{\hat G\{(\tilde{T}\wedge t)^-| A,L\}}\left\{N(dt)-\hat\Lambda_1^*(dt|A,L) Y(t)\right\}\right]\\
&\quad\quad+E\left\{\hat F_{1}^*(t_{0} | 1,L)-\hat F_{1}^*(t_{0} | 0,L)\right\}-E\left\{\Psi(\hat P^*_n)\right\}\\
\end{aligned}
$$
$$
\begin{aligned}
&=\int_{\mathcal{L}} \Sigma_{a=0,1}\int_0^{t_0} \frac{\{\mathbbm{1}(a=1)-\mathbbm{1}(a=0)\}\pi(a|l)}{\hat \pi(a| l)} \cdot \\
&\quad \quad\frac{\{1-\hat F_{1}^*\left(t_{0} | a, l\right)\}\{1- F_1 (t^- | a,l)\}G\{(\tilde T \wedge t)^-|a,l\}}{\{1-\hat F_1^* (t | a,l)\}\hat G\{(\tilde T \wedge t)^-|a,l\}}\left\{\Lambda_1(dt|a,l)-\hat\Lambda_1^*(dt|a,l)\right\}\mu(l)dl\\
&\quad\quad+E\left\{\hat F_{1}^*(t_{0} | 1,L)-\hat F_{1}^*(t_{0} | 0,L)\right\}-E\left\{\Psi(\hat P^*_n)\right\}\\
&\quad\quad(\textcolor{gray}{\text{Because }E\left\{f(Y)\mathbbm{1}(A=1)\right\}=E(f(Y)|A=1)P(A=1),}\\ &\quad\quad\quad\textcolor{gray}{\text{ where $f(Y)$ is a function of $Y$, and $A$ is a random variable}}\\
&=\int_{\mathcal{L}} \Sigma_{a=0,1} \int_0^{t_0}\left[\frac{\pi(a|  l)}{\hat \pi(a |  l)} \frac{G\{(\tilde T \wedge t)^-|a,l\}}{\hat G\{(\tilde T \wedge t)^-|a,l\}}\frac{\left\{1-\hat F_{1}^*\left(t_0|a,l\right)\right\}}{\left\{1-\hat F_1^*(t |  a, l)\right\} } \left\{\Lambda_{1}(d t |  a, l)-\hat \Lambda_1^*(d t |  a, l)\right\}\right.\\ 
&\quad\quad\quad \left.\left\{1- F_{1}(t^-|  a,l)\right\}\right]\{\mathbbm{1}(a=1)-\mathbbm{1}(a=0)\}\mu(l)dl\\
&\quad\quad+E\left\{\hat F_{1}^*(t_{0} | 1,L)-\hat F_{1}^*(t_{0} | 0,L)\right\}-E\left\{\Psi(\hat P^*_n)\right\}.
\end{aligned}
$$
Therefore, the second-order remainder
$$
\begin{aligned}
&R(\hat P^*_n,P)\\&=\Psi(\hat P^*_n)-\Psi(P)+E\left\{D(P^*_n)\right\}\\
&=\Psi(\hat P^*_n)-\Psi(P)\\&\quad+\int_{\mathcal{L}} \Sigma_{a=0,1} \int_0^{t_0}\left[\frac{\pi(a|  l)}{\hat \pi(a |  l)} \frac{G\{(\tilde T \wedge t)^-|a,l\}}{\hat G\{(\tilde T \wedge t)^-|a,l\}}\frac{\left\{1-\hat F_{1}^*\left(t_0|a,l\right)\right\}}{\left\{1-\hat F_1^*(t |  a, l)\right\} } \left\{\Lambda_{1}(d t |  a, l)-\hat \Lambda_1^*(d t |  a, l)\right\}\right.\\ 
&\quad\quad\quad \left.\left\{1- F_{1}(t^-|  a,l)\right\}\right]\{\mathbbm{1}(a=1)-\mathbbm{1}(a=0)\}\mu(l)dl\\
&\quad\quad+E\left\{\hat F_{1}^*(t_{0} | 1,L)-\hat F_{1}^*(t_{0} | 0,L)\right\}-E\left\{\Psi(\hat P^*_n)\right\}\\
&=\int_{\mathcal{L}} \Sigma_{a=0,1} \int_0^{t_0}\left[\frac{\pi(a|  l)}{\hat \pi(a |  l)} \frac{G\{(\tilde T \wedge t)^-|a,l\}}{\hat G\{(\tilde T \wedge t)^-|a,l\}}\frac{\left\{1-\hat F_{1}^*\left(t_0|a,l\right)\right\}}{\left\{1-\hat F_1^*(t |  a, l)\right\} } \left\{\Lambda_{1}(d t |  a, l)-\hat \Lambda_1^*(d t |  a, l)\right\}\right.\\ 
&\quad\quad\quad \left.\left\{1- F_{1}(t^-|  a,l)\right\}\right]\{\mathbbm{1}(a=1)-\mathbbm{1}(a=0)\}\mu(l)dl\\&\quad\quad +E\left\{\hat F_{1}^*(t_{0} | 1,L)-\hat F_{1}^*(t_{0} | 0,L)\right\}-\Psi(P)\\
\end{aligned}
$$
$$
\begin{aligned}
&=\int_{\mathcal{L}} \Sigma_{a=0,1} \int_0^{t_0}\left[\frac{\pi(a|  l)}{\hat \pi(a |  l)} \frac{G\{(\tilde T \wedge t)^-|a,l\}}{\hat G\{(\tilde T \wedge t)^-|a,l\}}\frac{\left\{1-\hat F_{1}^*\left(t_0|a,l\right)\right\}}{\left\{1-\hat F_1^*(t |  a, l)\right\} } \left\{\Lambda_{1}(d t |  a, l)-\hat \Lambda_1^*(d t |  a, l)\right\}\right.\\ 
&\quad\quad\quad \left.\left\{1- F_{1}(t^-|  a,l)\right\}\right]\{\mathbbm{1}(a=1)-\mathbbm{1}(a=0)\}\mu(l)dl\\
&\quad\quad\quad+\int_\mathcal{L}\Sigma_{a=0,1}\left\{\hat F_1^*(t_0|a,l)-F_1(t_0|a,l)\right\}\left\{\mathbbm{1}(a=1)-\mathbbm{1}(a=0)\right\}\mu(l)dl\\
&=\int_{\mathcal{L}} \Sigma_{a=0,1} \int_0^{t_0}\left[\frac{\pi(a|  l)}{\hat \pi(a |  l)} \frac{G\{(\tilde T \wedge t)^-|a,l\}}{\hat G\{(\tilde T \wedge t)^-|a,l\}}\frac{\left\{1-\hat F_{1}^*\left(t_0|a,l\right)\right\}}{\left\{1-\hat F_1^*(t |  a, l)\right\} } \left\{\Lambda_{1}(d t |  a, l)-\hat \Lambda_1^*(d t |  a, l)\right\}\right.\\ 
&\quad\quad\quad \left.\left\{1- F_{1}(t^-|  a,l)\right\}\right]\{\mathbbm{1}(a=1)-\mathbbm{1}(a=0)\}\mu(l)dl\\
&\quad\quad\quad+\int_\mathcal{L}\Sigma_{a=0,1}\left\{1-F_{1}(t^-|  a,l)\right\}\left\{\Lambda_{1}(dt|  a,l)-\hat\Lambda_{1}^*(dt|  a,l)\right\}\frac{\left\{1-\hat F_{1}^*(t_0|  a,l)\right\}}{\left\{1-\hat F_{1}^*(t|  a,l)\right\}}\\&\quad\quad\quad\left\{\mathbbm{1}(a=1)-\mathbbm{1}(a=0)\right\}\mu(l)dl \textcolor{gray}{\text{ \quad \quad by the Duhamel equation \citep{gill_lectures_1994},} }\\
&\textcolor{gray}{\quad  \hat F_{1}^*(t_0|  a,L)-F_{1}(t_0|  a,l)=-\int_0^{t_0} \left\{1-F_{1}(t^-|  a,l)\right\}\left\{\Lambda_{1}(dt|  a,l)-\hat\Lambda_{1}^*(dt|  a,l)\right\}\frac{\left\{1-\hat F_{1}^*\left(t_{0} |  a, l\right)\right\}}{\left\{1-\hat F_1^*(t |  a, l)\right\} } \text{}}\\
&=\int_{\mathcal{L}} \Sigma_{a=0,1} \int_0^{t_0}\left(\left[\frac{\pi(a|  l)}{\hat \pi(a |  l)} \frac{G\{(\tilde T \wedge t)^-|a,l\}}{\hat G\{(\tilde T \wedge t)^-|a,l\}}-1\right]\left\{\Lambda_{1}(d t |  a, l)-\hat \Lambda_1^*(d t |  a, l)\right\}\right.\\ 
&\quad\quad\quad \left.\frac{\left\{1-\hat F_{1}^*\left(t_0|a,l\right)\right\}}{\left\{1-\hat F_1^*(t |  a, l)\right\} } \left\{1- F_{1}(t^-|  a,l)\right\}\right)\{\mathbbm{1}(a=1)-\mathbbm{1}(a=0)\}\mu(l)dl\\
&=O_p\left(\Sigma_{a=0,1} \norm{\int_0^{t_0}\left[\pi(a|l)G\{(\tilde T \wedge t)^-|a,l\}-
\hat \pi(a |  l)\hat G\{(\tilde T \wedge t)^-|a,l\}\right]dt}\right.\cdot\\ &\quad\quad\quad \left.\norm{\int_0^{t_0}\left\{\Lambda_{1}(d t |  a, l)-\hat \Lambda_1^*(d t |  a, l)\right\}}\right),
\textcolor{gray}{\text{by the Cauchy-Schwarz inequality.}}\\
\end{aligned}
$$
\setlength{\emergencystretch}{0.5\textwidth}\par
The remainder is equal to $o_p(n^{-1/2})$ when either both the treatment and censoring mechanisms are consistently estimated or the CIF is consistently estimated, i.e., either $\norm{\{\pi(a| l)G(t| a,l)\}-\{\hat \pi(a| l){\hat G(t | a, l)}\}} = o_p(n^{-1/2})$, or $\norm{\Lambda_{1}(dt | a,l)-\hat \Lambda_1^*(dt | a,l)}= o_p(n^{-1/2})$. $R(\hat P_n^*,P)=o_p(n^{-1/2})$ can also be achieved if both $\hat \pi_0(a| l)\hat G(t | a, l)$ and $\hat \lambda_1^*(t|a,l)$ converge at a rate faster than $o_P(n^{-1/4})$.

\newpage
\section*{\textbf{Web Appendix E: Simulation Settings}}\label{app:c2sim}
\textbf{E.1 Simulation Study 1}\\
Results shown in Figure 1 is from this data setting. The baseline predictive covariates $V_1$, $V_2$, and the prognostic covariate $L_1$, $L_2$ and $L_3$ were independently generated from $Ber(0.5)$, and the other five prognostic covariates $L_4$, $L_5$, $L_6$, $L_7$ and $L_8$, were independently generated from $N(0,1)$. The binary treatment variable $A$ was generated from $Ber([1+\exp\{-(-0.2V_1-0.2V_2+1.5L_1+0.1L_2-0.1L_3+0.1L_4-0.1L_6)\}]^{-1})$. To generate the main event time $T_1$, we used a unit exponential mixture with the failure function taking the form: $F_1(t|A,V,L)=1-\{1-p(1-\exp(-t))\}^{\exp\{Y_1(A,V,L)\}}$, where $Y_1(A,V,L)=(0.6V_1-0.9L_1-0.1L_2+0.1L_3+0.1L_4-0.1L_5)+(-V_1+0.7V_2)A+0.5(A-0.5)$, $p=0.7$. The competing event time $T_2$ was generated from exponential distribution with mean $ [\exp\{Y_2(A,V, L)\}]^{-1}$, where $Y_2(A,V,L)=0.5Y_1(A,V,L)$. The event indicator $\Delta \sim 1+Ber\{(1-p)^{Y_1(A,V, L)}\}$. Finally, the censoring times $C$ were generated from exponential distribution with mean $[\lambda_0\exp\{0.2V_1-0.2V_2-0.2L_1-0.1L_2-0.1L_3+0.1L_4+0.1L_7\}]^{-1}$, resulting in approximately 51\% for the main event, 28\% for the competing event, and 21\% for noninformative censoring. The interaction terms between treatment $A$ and covariates $V_1$ and $V_2$ in $Y_1$ introduced the treatment effect heterogeneity across the simulation scenarios.

To investigate the double robustness and asymptotic properties of our proposed estimator, we included scenarios with correctly specified and misspecified models for the outcome, treatment and censoring. In the misspecified outcome models, $L_1$, $L_2$ and $L_5$ were deliberately excluded, with additional covariates $L_6$, $L_7$ and $L_8$ introduced in the subdistribution hazard model. The misspecified treatment models excluded $L_1$, $L_2$ and $L_6$, and introduced additional covariates, $L_5$, $L_7$ and $L_8$. Similarly, for the misspcification in censoring models,  we deliberately omitted $L_1$, $L_2$ and $L_7$, and introduced extra covariates $L_5$, $L_6$ and $L_8$. To provide a benchmark, we implemented existing meta-algorithms for subgroup ATE, specifically the S-learner and T-learner as described in \citep{kunzel_meta-learners_2019}, and compared their performance to our proposed TMLE framework. The S-learner utilizes a single model to predict the outcome, with treatment as a covariate and including its interaction terms with covariates, and calculates the difference in predictions between different treatment assignments. Conversely, the T-learner uses two separate models, one for treated and one for untreated individuals, and subsequently computes the difference in predictions. The S-learner and T-learner were also incorporated in the first step in our TMLE framework to generate the initial estimates of the outcome. Simulations were also conducted using one-step estimator for comparison, which is another doubly robust estimator based on our derived EIF, and details regarding one-step estimator were presented in Web Appendix F. 

\vspace{2em}
\textbf{E.2 Simulation Study 2}\\
In addition to evaluating the properties of our proposed estimator and comparison with existing methods, we estimated subgroup ATE by incorporating various combinations of models in our TMLE framework, to investigate the performance in data with many noise variables that are independent of the subgroup ATE. For the outcome model, we employed subdistribution hazards model or LASSO-regularized subdistribution hazards model, implemented using the R package \texttt{fastcmprsk} \citep{kawaguchi_fast_2020}. The treatment model was built using logistic regression or a super learner that encompasses generalized linear models, generalized additive models, and LASSO-regularized logistic regression using the R package \texttt{SuperLearner} \citep{van2007super}. The censoring model was built using Cox model, or a super learner that encompasses Cox model, exponential model, log-logistic model, and LASSO-regularized Cox model using the R package \texttt{survSuperLearner} \citep{golmakani2020super}. We applied these methods in another set of simulated data to mimic the a scenario with a large amount of covariates irrelevant for estimating subgroup ATE. Twenty additional variables $L_9, L_{10}, ..., L_{28}\sim N(0,1)$, which were independent of all event time, censoring time and treatment probabilities, were added to each simulated dataset and included in the models to evaluate the performance in variable selection.

\vspace{2em}
\textbf{E.3 Methods Evaluation}\\
Our simulations included adjustments to the sample size ($n$) of 800, 1,500, and 3,000, and the time points of interests at the 25th, 50th and 75th percentile event time. We conducted $B=500$ simulations under the aforementioned settings.  Estimates are generated for each one of the four subgroups defined by $V_1$ and $V_2$: group $m=1$ with $V_1=0$ and $V_2=0$, group $m=2$ with $V_1=0$ and $V_2=1$, group $m=3$ with $V_1=1$ and $V_2=0$, and group $m=4$ with $V_1=1$ and $V_2=1$. Bias and root-mean-square error (RMSE) were computed for each subgroup ($m=1,2,3,4$) defined by $V_1$ and $V_2$ as $\text{Bias}_m(t_0)=\Psi_m(\hat{P}^*_n)-\Psi_m(P) \text{ and RMSE }_m(t_0)=\sqrt{\frac{1}{B}\sum^B_{b=1}\{\Psi_m(\hat{P}^*_n)-\Psi_m(P)\}^2}$,
where $\Psi_m(\hat{P}^*_n)$ and $\Psi_m(P)$ represent the estimated and true subgroup ATE, respectively for subgroup $m$ at time $t_0$. Additionally, we calculate the 95\% confidence interval (CI) for the estimated subgroup ATE and the proportion of true subgroup ATE covered by the 95\% CI. Hypothesis tests are conducted to make inference about whether the subgroup ATE in each subgroup is significantly different from 0.

\newpage
\section*{\textbf{Web Appendix F: One-step estimator}}
In addition to S-learner and T-learner, we also compare our method with another doubly robust estimator, the one-step estimator (also known as the bias-corrected estimator). This estimator has been used for continuous outcome, and to compare it with TMLE, we adapted it by incorporating our derived EIF in (7). Instead of finding the targeted distribution $\hat P_n^*$ by solving the efficient influence function equation (4) as in TMLE, the one-step estimator calculate and substract the empirical average of the estimated EIF to the plug-in estimator, which is built by plugging in initial estimator:
\begin{equation}
\begin{aligned}
 \Psi_{os}(\hat P_n)=\Psi_{pi}(\hat P_n)+\mathbb{P}_n\{D(\hat P_n)\},
\end{aligned}
\end{equation}
where $\Psi_{os}(\hat P_n)$ is the one-step estimator, $\Psi_{pi}(\hat P_n)$ is the plug-in estimator, and $D(P)$ is our derived EIF in (7). S-learner and T-learner can be incorporated into the one-step estimator for the outcome model.

The one-step estimator should share similar asymptotic properties with TMLE, but because it is not a substitution estimator, the estimated value may go beyond the range (-1,1), especially when the positivity assumption is nearly violated. In the simulation studies presented in Web Table 1-4, we could see that in large sample (N=3000), one-step estimator has similar performance as TMLE methods. When the sample size decreases, especially in N=800, we observe higher RMSE using one-step estimators. This could be caused by the nearly violated positivity assumption in subgroups when the sample size is small, and some estimated value of one-step estimators could be extreme or beyond the range.

\newpage
\section*{\textbf{Web Appendix G: Additional Simulation Results}}
\textbf{Web Table 1.} Performance of subgroup ATE estimations at multiple follow-up time point for the selected stratum $V_1=0$ and $V_2=0$, two binary predictive variables (N=3000).$^a$ \label{V00T25}\\
\small
\rowcolors{3}{gray!8}{white}
\begin{tabular}{l c c c c c c c c c c}
\toprule
&  \multicolumn{2}{c}{ $\begin{array}{c}\textbf {Scenario 1$^c$} \\
\text {Outcome } \checkmark\\\text {Treatment }\checkmark \\\text {Censoring }\checkmark
\end{array}$} & \multicolumn{2}{c}{$\begin{array}{c} \textbf {Scenario 2$^c$}  \\
\text {Outcome } \times\\\text {Treatment } \checkmark \\\text {Censoring } \checkmark
\end{array}$} & \multicolumn{2}{c}{$\begin{array}{c}\textbf {Scenario 3$^c$} \\
\text {Outcome } \checkmark \\\text {Treatment }  \times\\\text {Censoring }\checkmark \end{array}$} 
& \multicolumn{2}{c}{$\begin{array}{c}\textbf {Scenario 4$^c$} \\
\text {Outcome }  \checkmark\\\text {Treatment }  \checkmark\\\text {Censoring }  \times\end{array}$} 
& \multicolumn{2}{c}{$\begin{array}{c}\textbf {Scenario 5$^c$} \\
\text {Outcome }  \checkmark\\\text {Treatment }  \times\\\text {Censoring }  \times
\end{array}$} \\
\cmidrule(lr){2-3}\cmidrule(lr){4-5}\cmidrule(lr){6-7}\cmidrule(lr){8-9}\cmidrule(lr){10-11}
$\begin{array}{l}
\textbf{Algorithms$^b$}\end{array}$ & 
\text {Bias$^d$}& RMSE & \text {Bias$^d$}& RMSE &\text {Bias$^d$}& RMSE &\text {Bias$^d$}& RMSE &\text {Bias$^d$}& RMSE \\
\hline
\rowcolor{gray!30}
\multicolumn{11}{c}{At the 25th percentile follow-up time point} \\
S-learner & 0.000 & 0.010 & -0.032 & 0.034 & ~ & ~ & ~ & ~ & ~ & ~  \\ 
T-learner & -0.001 & 0.013 & -0.03 & 0.033 & ~ & ~ & ~ & ~ & ~ & ~  \\ 
TMLE+S & 0.001 & 0.025 & 0.001 & 0.025 & 0.001 & 0.023 & 0.001 & 0.025 & 0.001 & 0.023  \\ 
TMLE+T & 0.001 & 0.025 & 0.000 & 0.025 & 0.001 & 0.023 & 0.001 & 0.025 & 0.001 & 0.023  \\ 
OS+S & 0.001 & 0.025 & 0.001 & 0.025 & 0.001 & 0.024 & 0.001 & 0.025 & 0.001 & 0.024  \\ 
OS+T & 0.001 & 0.025 & 0.000 & 0.025 & 0.001 & 0.023 & 0.001 & 0.025 & 0.001 & 0.023  \\ 
\hline
\rowcolor{gray!30}
\multicolumn{11}{c}{At the 50th percentile follow-up time point} \\
S-learner & 0.000 & 0.018 & -0.062 & 0.065 & ~ & ~ & ~ & ~ & ~ & ~  \\ 
T-learner & -0.001 & 0.021 & -0.059 & 0.063 & ~ & ~ & ~ & ~ & ~ & ~  \\ 
TMLE+S & 0.002 & 0.034 & 0.002 & 0.034 & 0.001 & 0.033 & 0.002 & 0.034 & 0.001 & 0.033  \\ 
TMLE+T & 0.002 & 0.034 & 0.001 & 0.034 & 0.001 & 0.033 & 0.002 & 0.034 & 0.001 & 0.033  \\ 
OS+S & 0.002 & 0.034 & 0.002 & 0.034 & 0.001 & 0.034 & 0.002 & 0.034 & 0.001 & 0.034  \\ 
OS+T & 0.002 & 0.034 & 0.001 & 0.034 & 0.001 & 0.033 & 0.002 & 0.034 & 0.001 & 0.033  \\ \hline 
\rowcolor{gray!30}
\multicolumn{11}{c}{At the 75th percentile follow-up time point} \\
S-learner & 0.000 & 0.026 & -0.087 & 0.092 & ~ & ~ & ~ & ~ & ~ & ~  \\ 
T-learner & -0.001 & 0.028 & -0.087 & 0.092 & ~ & ~ & ~ & ~ & ~ & ~  \\ 
TMLE+S & 0.000 & 0.040 & -0.001 & 0.041 & -0.001 & 0.038 & 0.000 & 0.040 & -0.001 & 0.038  \\ 
TMLE+T & 0.000 & 0.040 & -0.002 & 0.041 & 0.000 & 0.039 & 0.000 & 0.040 & 0.000 & 0.039  \\ 
OS+S & 0.000 & 0.040& -0.001 & 0.041 & -0.001 & 0.039 & 0.000 & 0.040 & -0.001 & 0.039  \\ 
OS+T & 0.000 & 0.040 & -0.001 & 0.041 & -0.001 & 0.039 & 0.0000 & 0.040 & -0.001 & 0.039 \\ 
\bottomrule
\end{tabular}\\
\footnotesize{
\RaggedRight {$^a$ Only the subgroup $V_1=0$ and $V_2=0$ is selected to be reported in this table due to the limited space and similar performance in other subgroups. \\
$^b$Abbreviations: S: S-learner; T: T-learner; OS: One-step estimator.\\
$^c$$\checkmark$ and $\times$ indicate that the corresponding models were correctly specified and misspecified, respectively. \\
$^d$Mean bias and RMSE are calculated across 500 Monte-Carlo repetitions.}}

\newpage
\normalsize
\RaggedRight {\textbf{Web Table 2.} Performance of subgroup ATE estimations at multiple follow-up time point for the selected stratum $V_1=0$ and $V_2=0$, two binary predictive variables (N=1500).$^a$ }\label{V00T25}\\
\small
\rowcolors{3}{gray!8}{white}
\begin{tabular}{l c c c c c c c c c c}
\toprule
&  \multicolumn{2}{c}{ $\begin{array}{c}\textbf {Scenario 1$^c$} \\
\text {Outcome } \checkmark\\\text {Treatment }\checkmark \\\text {Censoring }\checkmark
\end{array}$} & \multicolumn{2}{c}{$\begin{array}{c} \textbf {Scenario 2$^c$}  \\
\text {Outcome } \times\\\text {Treatment } \checkmark \\\text {Censoring } \checkmark
\end{array}$} & \multicolumn{2}{c}{$\begin{array}{c}\textbf {Scenario 3$^c$} \\
\text {Outcome } \checkmark \\\text {Treatment }  \times\\\text {Censoring }\checkmark \end{array}$} 
& \multicolumn{2}{c}{$\begin{array}{c}\textbf {Scenario 4$^c$} \\
\text {Outcome }  \checkmark\\\text {Treatment }  \checkmark\\\text {Censoring }  \times\end{array}$} 
& \multicolumn{2}{c}{$\begin{array}{c}\textbf {Scenario 5$^c$} \\
\text {Outcome }  \checkmark\\\text {Treatment }  \times\\\text {Censoring }  \times
\end{array}$} \\
\cmidrule(lr){2-3}\cmidrule(lr){4-5}\cmidrule(lr){6-7}\cmidrule(lr){8-9}\cmidrule(lr){10-11}
$\begin{array}{l}
\textbf{Algorithms$^b$}\end{array}$ & 
\text {Bias$^d$}& RMSE & \text {Bias$^d$}& RMSE &\text {Bias$^d$}& RMSE &\text {Bias$^d$}& RMSE &\text {Bias$^d$}& RMSE \\
\hline
\rowcolor{gray!30}
\multicolumn{11}{c}{At the 25th percentile follow-up time point} \\
S-learner & -0.001 & 0.014 & -0.033 & 0.035 & ~ & ~ & ~ & ~ & ~ & ~ \\ 
T-learner & -0.001 & 0.018 & -0.030 & 0.035 & ~ & ~ & ~ & ~ & ~ & ~ \\ 
TMLE+S & -0.001 & 0.034 & -0.001 & 0.034 & -0.001 & 0.032 & -0.001 & 0.034 & -0.001 & 0.032  \\ 
TMLE+T & -0.001 & 0.034 & -0.002 & 0.034 & -0.001 & 0.031 & -0.001 & 0.034 & -0.001 & 0.031  \\ 
OS+S & -0.001 & 0.034 & -0.001 & 0.034 & -0.001 & 0.033 & -0.001 & 0.034 & -0.001 & 0.033  \\ 
OS+T & -0.001 & 0.034 & -0.002 & 0.034 & -0.001 & 0.031 & -0.001 & 0.034 & -0.001 & 0.031  \\ \hline
\rowcolor{gray!30}
\multicolumn{11}{c}{At the 50th percentile follow-up time point} \\
S-learner & -0.001 & 0.026 & -0.063 & 0.068 & ~ & ~ & ~ & ~ & ~ & ~ \\ 
T-learner & -0.002 & 0.029 & -0.060 & 0.067 & ~ & ~ & ~ & ~ & ~ & ~ \\ 
TMLE+S & 0.001 & 0.049 & 0.001 & 0.049 & 0.001 & 0.047 & 0.002 & 0.049 & 0.001 & 0.047  \\ 
TMLE+T & 0.002 & 0.049 & 0.000 & 0.049 & 0.001 & 0.047 & 0.002 & 0.049 & 0.001 & 0.047  \\ 
OS+S & 0.002 & 0.049 & 0.001 & 0.049 & 0.001 & 0.048 & 0.002 & 0.049 & 0.001 & 0.048  \\ 
OS+T & 0.002 & 0.049 & 0.000 & 0.049 & 0.001 & 0.047 & 0.002 & 0.049 & 0.001 & 0.047  \\
\hline 
\rowcolor{gray!30}
\multicolumn{11}{c}{At the 75th percentile follow-up time point} \\
S-learner & -0.001 & 0.037 & -0.088 & 0.096 & ~ & ~ & ~ & ~ & ~ & ~ \\ 
T-learner & -0.002 & 0.039 & -0.088 & 0.097 & ~ & ~ & ~ & ~ & ~ & ~ \\ 
TMLE+S & 0.001 & 0.056 & -0.001 & 0.056 & -0.001 & 0.053 & 0.001 & 0.056 & -0.001 & 0.053  \\ 
TMLE+T & 0.001 & 0.056 & -0.002 & 0.056 & -0.001 & 0.054 & 0.001 & 0.056 & -0.001 & 0.054  \\ 
OS+S & 0.001 & 0.057 & 0.000 & 0.056 & -0.001 & 0.054 & 0.001 & 0.057 & -0.001 & 0.054  \\ 
OS+T & 0.001 & 0.056 & -0.001 & 0.056 & -0.001 & 0.054 & 0.001 & 0.056 & -0.001 & 0.054 \\
\bottomrule
\end{tabular}\\
\footnotesize{
\RaggedRight {$^a$ Only the subgroup $V_1=0$ and $V_2=0$ is selected to be reported in this table due to the limited space and similar performance in other subgroups. \\
$^b$Abbreviations: S: S-learner; T: T-learner; OS: One-step estimator.\\
$^c$$\checkmark$ and $\times$ indicate that the corresponding models were correctly specified and misspecified, respectively. \\
$^d$Mean bias and RMSE are calculated across 500 Monte-Carlo repetitions.}}

\newpage
\normalsize
\RaggedRight {\textbf{Web Table 3.} Performance of subgroup ATE estimations at multiple follow-up time point for the selected stratum $V_1=0$ and $V_2=0$, two binary predictive variables (N=800).$^a$} \label{V00T25}\\
\small
\rowcolors{3}{gray!8}{white}
\begin{tabular}{l c c c c c c c c c c}
\toprule
&  \multicolumn{2}{c}{ $\begin{array}{c}\textbf {Scenario 1$^c$} \\
\text {Outcome } \checkmark\\\text {Treatment }\checkmark \\\text {Censoring }\checkmark
\end{array}$} & \multicolumn{2}{c}{$\begin{array}{c} \textbf {Scenario 2$^c$}  \\
\text {Outcome } \times\\\text {Treatment } \checkmark \\\text {Censoring } \checkmark
\end{array}$} & \multicolumn{2}{c}{$\begin{array}{c}\textbf {Scenario 3$^c$} \\
\text {Outcome } \checkmark \\\text {Treatment }  \times\\\text {Censoring }\checkmark \end{array}$} 
& \multicolumn{2}{c}{$\begin{array}{c}\textbf {Scenario 4$^c$} \\
\text {Outcome }  \checkmark\\\text {Treatment }  \checkmark\\\text {Censoring }  \times\end{array}$} 
& \multicolumn{2}{c}{$\begin{array}{c}\textbf {Scenario 5$^c$} \\
\text {Outcome }  \checkmark\\\text {Treatment }  \times\\\text {Censoring }  \times
\end{array}$} \\
\cmidrule(lr){2-3}\cmidrule(lr){4-5}\cmidrule(lr){6-7}\cmidrule(lr){8-9}\cmidrule(lr){10-11}
$\begin{array}{l}
\textbf{Algorithms$^b$}\end{array}$ & 
\text {Bias$^d$}& RMSE & \text {Bias$^d$}& RMSE &\text {Bias$^d$}& RMSE &\text {Bias$^d$}& RMSE &\text {Bias$^d$}& RMSE \\
\hline
\rowcolor{gray!30}
\multicolumn{11}{c}{At the 25th percentile follow-up time point} \\
S-learner & 0 & 0.018 & -0.033 & 0.038 & ~ & ~ & ~ & ~ & ~ &   \\ 
T-learner & -0.001 & 0.024 & -0.03 & 0.039 & ~ & ~ & ~ & ~ & ~ &   \\ 
TMLE+S & -0.002 & 0.049 & -0.002 & 0.049 & -0.001 & 0.047 & -0.002 & 0.049 & -0.001 & 0.047  \\ 
TMLE+T & -0.002 & 0.049 & -0.003 & 0.049 & -0.002 & 0.047 & -0.002 & 0.049 & -0.002 & 0.047  \\ 
OS+S & -0.001 & 0.049 & -0.002 & 0.049 & -0.002 & 0.049 & -0.001 & 0.049 & -0.002 & 0.049  \\ 
OS+T & -0.001 & 0.049 & -0.003 & 0.049 & -0.002 & 0.047 & -0.001 & 0.049 & -0.002 & 0.047  \\ \hline
\rowcolor{gray!30}
\multicolumn{11}{c}{At the 50th percentile follow-up time point} \\
S-learner & 0 & 0.035 & -0.063 & 0.073 & ~ & ~ & ~ & ~ & ~ &   \\ 
T-learner & -0.003 & 0.039 & -0.061 & 0.074 & ~ & ~ & ~ & ~ & ~ &   \\ 
TMLE+S & -0.002 & 0.065 & -0.003 & 0.066 & -0.002 & 0.064 & -0.002 & 0.066 & -0.002 & 0.064  \\ 
TMLE+T & -0.002 & 0.065 & -0.005 & 0.066 & -0.003 & 0.064 & -0.002 & 0.066 & -0.003 & 0.064  \\ 
OS+S & -0.002 & 0.066 & -0.003 & 0.066 & -0.003 & 0.066 & -0.002 & 0.066 & -0.003 & 0.066  \\ 
OS+T & -0.002 & 0.066 & -0.005 & 0.066 & -0.003 & 0.064 & -0.002 & 0.066 & -0.003 & 0.064  \\ \hline 
\rowcolor{gray!30}
\multicolumn{11}{c}{At the 75th percentile follow-up time point} \\
S-learner & -0.001 & 0.048 & -0.089 & 0.103 & ~ & ~ & ~ & ~ & ~ &   \\ 
T-learner & -0.004 & 0.052 & -0.09 & 0.106 & ~ & ~ & ~ & ~ & ~ &   \\ 
TMLE+S & -0.005 & 0.084 & -0.006 & 0.084 & -0.005 & 0.077 & -0.005 & 0.084 & -0.005 & 0.077  \\ 
TMLE+T & -0.004 & 0.083 & -0.01 & 0.085 & -0.006 & 0.079 & -0.004 & 0.083 & -0.006 & 0.079  \\ 
OS+S & -0.005 & 0.084 & -0.006 & 0.084 & -0.005 & 0.079 & -0.005 & 0.084 & -0.005 & 0.079  \\ 
OS+T & -0.004 & 0.083 & -0.009 & 0.085 & -0.006 & 0.079 & -0.004 & 0.083 & -0.006 & 0.079 \\ 
\bottomrule
\end{tabular}\\
\footnotesize{
\RaggedRight {$^a$ Only the subgroup $V_1=0$ and $V_2=0$ is selected to be reported in this table due to the limited space and similar performance in other subgroups. \\
$^b$Abbreviations: S: S-learner; T: T-learner; OS: One-step estimator.\\
$^c$$\checkmark$ and $\times$ indicate that the corresponding models were correctly specified and misspecified, respectively. \\
$^d$Mean bias and RMSE are calculated across 500 Monte-Carlo repetitions.}}

\newpage
\normalsize
\textbf{Web Table 4.} Performance of subgroup ATE estimations at the 50th percentile follow-up time point using data with noise variables. \label{V00T25}\\
\footnotesize
\rowcolors{3}{gray!8}{white}
\setlength\extrarowheight{-2.4pt}
\begin{tabular}{l c c c c c c c c}
\toprule
&  \multicolumn{2}{c}{ $\begin{array}{c}\textbf {Subgroup 1}\\ \text {$V_1=0$, $V_2=0$}\end{array}$} 
&  \multicolumn{2}{c}{ $\begin{array}{c}\textbf {Subgroup 2}\\ \text {$V_1=0$, $V_2=1$}\end{array}$} 
&  \multicolumn{2}{c}{ $\begin{array}{c}\textbf {Subgroup 3}\\ \text {$V_1=1$, $V_2=0$}\end{array}$} 
&  \multicolumn{2}{c}{ $\begin{array}{c}\textbf {Subgroup 4}\\ \text {$V_1=1$, $V_2=1$}\end{array}$} 
\\
\cmidrule(lr){2-3}\cmidrule(lr){4-5}\cmidrule(lr){6-7}\cmidrule(lr){8-9}
$\begin{array}{l}
\textbf{Algorithms$^a$}\end{array}$ & 
\text {Bias$^b$}& RMSE & \text {Bias$^b$}& RMSE &\text {Bias$^b$}& RMSE &\text {Bias$^b$}& RMSE \\
\hline
\rowcolor{gray!30}
\multicolumn{9}{c}{N=3000} \\
TMLE+S & -0.003 & 0.036 & 0.001 & 0.036 & 0.000 & 0.037 & -0.001 & 0.038  \\ 
TMLE+T & -0.004 & 0.036 & 0.000 & 0.036 & 0.000 & 0.037 & -0.002 & 0.038  \\ 
TMLE+S+LASSO & -0.003 & 0.036 & 0.001 & 0.035 & 0.000 & 0.037 & -0.001 & 0.038  \\ 
TMLE+T+LASSO & -0.003 & 0.036 & 0.001 & 0.036 & 0.000 & 0.037 & -0.001 & 0.038  \\ 
TMLE+S+SL & -0.002 & 0.035 & 0.001 & 0.035 & 0.000 & 0.037 & -0.001 & 0.037  \\ 
TMLE+T+SL & -0.003 & 0.035 & 0.000 & 0.035 & 0.001 & 0.036 & -0.001 & 0.037  \\ 
TMLE+S+SL+LASSO & -0.002 & 0.035 & 0.001 & 0.035 & 0.000 & 0.037 & -0.001 & 0.037  \\ 
TMLE+T+TL+LASSO & -0.003 & 0.034 & 0.001 & 0.034 & 0.001 & 0.036 & -0.001 & 0.037  \\ 
OS+S & -0.003 & 0.036 & 0.001 & 0.036 & 0.000 & 0.037 & -0.001 & 0.038  \\ 
OS+T & -0.004 & 0.036 & 0.000 & 0.036 & 0.000 & 0.037 & -0.002 & 0.038  \\ 
OS+S+LASSO & -0.003 & 0.036 & 0.001 & 0.036 & 0.000 & 0.037 & -0.001 & 0.038  \\ 
OS+T+LASSO & -0.003 & 0.036 & 0.001 & 0.036 & 0.000 & 0.037 & -0.001 & 0.038  \\ 
OS+S+SL & -0.002 & 0.034 & 0.001 & 0.034 & 0.000 & 0.035 & -0.001 & 0.036  \\ 
OS+T+SL & -0.003 & 0.034 & 0.000 & 0.035 & 0.001 & 0.036 & -0.001 & 0.037  \\ 
OS+S+SL+LASSO & -0.002 & 0.034 & 0.001 & 0.034 & 0.001 & 0.035 & -0.002 & 0.036  \\ 
OS+T+SL+LASSO & -0.003 & 0.034 & 0.001 & 0.034 & 0.001 & 0.036 & -0.001 & 0.037 \\ 
\hline
\rowcolor{gray!30}
\multicolumn{9}{c}{N=800} \\
TMLE+S & 0.000 & 0.086 & -0.006 & 0.080 & -0.008 & 0.090 & -0.004 & 0.087  \\ 
TMLE+T & -0.027 & 0.089 & -0.027 & 0.081 & -0.017 & 0.088 & -0.013 & 0.084  \\ 
TMLE+S+LASSO & 0.000 & 0.086 & -0.006 & 0.080 & -0.009 & 0.090 & -0.004 & 0.087  \\ 
TMLE+T+LASSO & -0.019 & 0.087 & -0.016 & 0.078 & -0.013 & 0.088 & -0.009 & 0.083  \\ 
TMLE+S+SL & 0.002 & 0.074 & -0.002 & 0.068 & -0.002 & 0.079 & -0.002 & 0.078  \\ 
TMLE+T+SL & -0.027 & 0.085 & -0.025 & 0.073 & -0.012 & 0.080 & -0.010 & 0.077  \\ 
TMLE+S+SL+LASSO & 0.000 & 0.073 & -0.003 & 0.067 & -0.004 & 0.076 & -0.003 & 0.076  \\ 
TMLE+T+TL+LASSO & -0.019 & 0.081 & -0.014 & 0.069 & -0.010 & 0.077 & -0.007 & 0.075  \\ 
OS+S & 0.005 & 0.087 & -0.003 & 0.082 & -0.006 & 0.096 & -0.003 & 0.086  \\ 
OS+T & -0.029 & 0.092 & -0.027 & 0.085 & -0.018 & 0.093 & -0.012 & 0.088  \\ 
OS+S+LASSO & 0.004 & 0.088 & -0.004 & 0.083 & -0.006 & 0.100 & -0.004 & 0.086  \\ 
OS+T+LASSO & -0.020 & 0.089 & -0.015 & 0.082 & -0.014 & 0.094 & -0.007 & 0.088  \\ 
OS+S+SL & 0.003 & 0.068 & 0.000 & 0.065 & -0.003 & 0.075 & -0.002 & 0.074  \\ 
OS+T+SL & -0.031 & 0.084 & -0.026 & 0.073 & -0.012 & 0.080 & -0.010 & 0.076  \\ 
OS+S+SL+LASSO & -0.001 & 0.068 & -0.002 & 0.064 & -0.002 & 0.071 & -0.005 & 0.071  \\ 
OS+T+SL+LASSO & -0.023 & 0.081 & -0.015 & 0.068 & -0.010 & 0.078 & -0.007 & 0.075 \\ 
\hline 
\rowcolor{gray!30}
\bottomrule
\end{tabular}\\
\footnotesize{
\RaggedRight {$^a$Abbreviations: S: S-learner; T: T-learner; OS: One-step estimator; 
LASSO: initial estimates of CIF obtained through LASSO-penalized subdistribution hazards models; SL: super learner, which encompasses generalized linear models, generalized additive models, and LASSO-regularized logistic regression for treatment, and incorporates Cox models, or a super learner that encompasses Cox models, exponential models, log-logistic models, and LASSO-regularized Cox models for censoring.\\
$^b$Mean bias and RMSE are calculated across 500 Monte-Carlo repetitions.}}

\newpage
\normalsize
\section*{\textbf{Web Appendix H: Additional Analysis Results of Sepsis Dataset}}
\textbf{Web Table 5. }{Baseline characteristics of patients, grouped by untreated and treated by steroids$^a$.}\label{c2t:des}\\
\vspace{1em}
\begin{tabular}{p{6cm} ccc}
\hline
$\begin{array}{l}\textbf{Baseline} \\ \textbf{Variable} \end{array}$ & $\begin{array}{c}\textbf {No Steroids} \\ \textbf{(N=2,610)} \end{array}$ & 
$\begin{array}{c}\textbf {Steroids} \\ \textbf{(N=635)} \end{array}$ & \textbf{P-value$^a$} \\ 
\hline
\textbf{Age (Year)} & 63.8 (16.6) & 62.9 (14.8) & 0.193   \\
\textbf{Gender - Female} & 1063 (40.7\%)     & 313 (49.3\%)   & 0.04    \\
\textbf{Race} &         &               &                   \\
\quad White    & 2081 (79.7\%)         & 505 (79.5\%)       & 0.538             \\
\quad Black    & 289 (11.1\%)          & 78 (12.3\%)        &                   \\
\quad Other   & 240 (9.2\%)           & 52 (8.2\%)         &                   \\
\textbf{BMI}     & 29.9 (8.92)          & 29.4 (9.14)       & 0.158             \\
\textbf{Surgery - Yes}        & 1003 (38.4\%)         & 190 (29.9\%)       & $<$0.001            \\
\textbf{Sequential Organ Failure Assessment (SOFA)} & 10.1 (3.26)     & 11.7 (3.85)   & $<$0.001     \\
\textbf{Charlson Comorbidity Index (CCI)}           & 2.89 (2.33)          & 3.22 (2.37)       & 0.001     \\
\textbf{Glasgow Coma Scale (GCS)}                   & 7.07 (3.88)          & 7.08 (4.11)       & 0.97   \\
\textbf{Serum Lactate}                              & 4.82 (4.04)          & 6.83 (5.82)       & $<$0.001   \\ \hline
\end{tabular}
\\
\RaggedRight {$^a$ For continuous variables, we report mean (SD) and compare using t-test. For categorical variables, we report frequency (\%) and compare using Chi-squared test. }\\

\centerline{\includegraphics[width=0.9\textwidth]{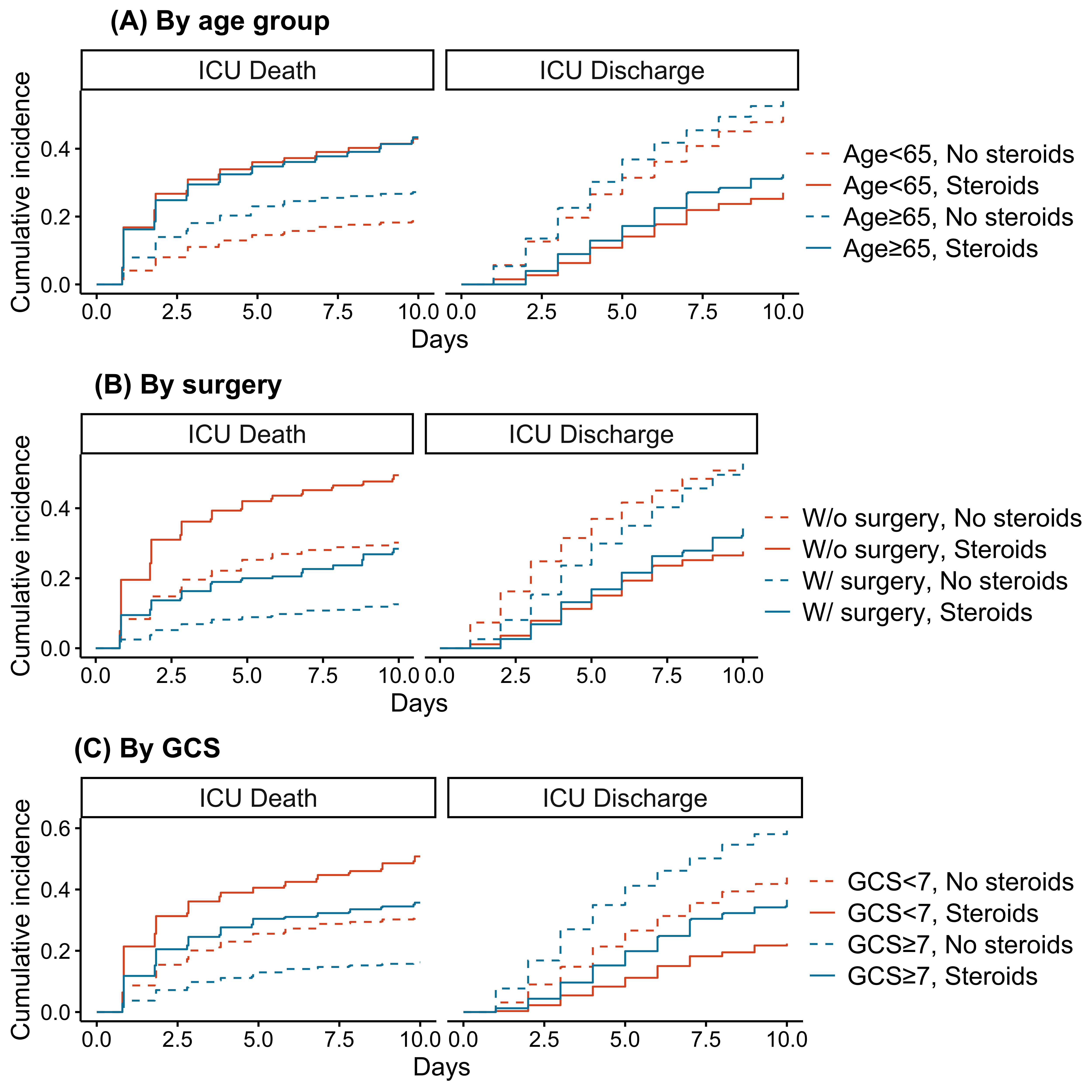}}
\textbf{Web Figure 1. }{Cumulative incidence plot of death and discharge from ICU within 10 days post-ICU admission, grouped by treatment assignment and (A) age group, (B) surgery status, and (C) Glasgow Coma Scale (GCS).}
\label{c2f:cif}

\centerline{\includegraphics[width=0.7\textwidth]{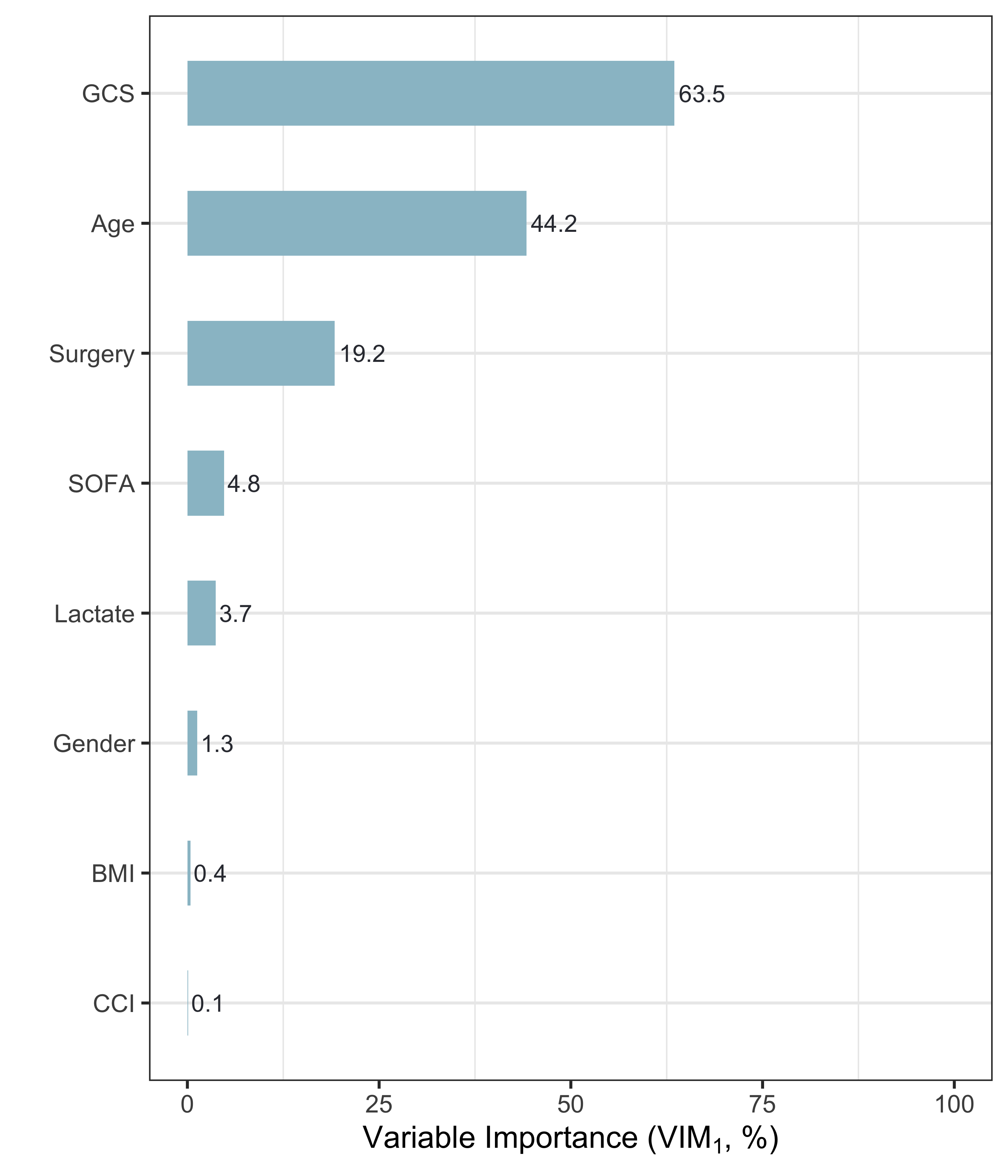}}
\textbf{Web Figure 2. }{Variable importance for predictive and prognostic variables, estimated by $\text{VIM}_1$, represents the difference in empirical variance of the treatment effect when each variable is ignored during analysis. }\label{fig:vim_all }

\backmatter

 \bibliographystyle{biom} 
\bibliography{ref.bib}

\label{lastpage}